\documentclass[prd, aps, preprint, floats, nofootinbib, tightenlines, showpacs]{revtex4} 

\usepackage{epsfig,amsfonts}
  
\begin{document} 

\pagestyle{empty}
\preprint{
\noindent
\begin{minipage}[t]{3in}
\begin{flushleft}
\end{flushleft}
 \end{minipage}
 \hfill
 \begin{minipage}[t]{3in}
\begin{flushright}
\end{flushright}
\end{minipage}
}

\title{Composite gauge-bosons made of fermions}

\author{Mahiko Suzuki}

\affiliation{
Lawrence Berkeley National Laboratory and Department of Physics\\
University of California, Berkeley, California 94720
}

\date{\today}

\begin{abstract}
We construct a class of Abelian and non-Abelian local gauge theories  
that consist only of matter fields of fermions. The Lagrangian is 
local and does not contain an auxiliary vector field nor a subsidiary 
condition on the matter fields. It does not involve an extra dimension
nor supersymmetry.  This Lagrangian can be extended to 
non-Abelian gauge symmetry only in the case of SU(2) doublet matter 
fields. We carry out explicit diagrammatic computation in the leading 
1/N order to show that massless spin-one bound states appear with the 
correct gauge coupling. Our diagram calculation exposes the dynamical 
features that cannot be seen in the formal auxiliary vector-field 
method.  For instance, it shows that the $s$-wave fermion-antifermion 
interaction in the $^3S_1$ channel ($\overline{\psi}\gamma_{\mu}\psi$) 
alone cannot form the bound gauge bosons; the fermion-antifermion pairs 
must couple to the $d$-wave state too. One feature common to our
class of Lagrangian is that the Noether current does not exist.
Therefore it evades possible conflict with the no-go theorem of 
Weinberg and Witten on formation of the non-Abelian gauge bosons. 

\end{abstract} 

  (The version to appear in Phys. Rev. D)

\vskip 1cm

\maketitle

\pagestyle{plain}

\setcounter{footnote}{0}

\section{Introduction}
  The U(1) gauge theory normally consists of a gauge field and matter fields. 
The Lagrangian is invariant under the simultaneous gauge transformation of 
the gauge field and the matter fields. After this was generalized to non-Abelian 
group\cite{YM}, we learned that the non-Abelian extension underlies dynamics 
of the fundamental particles.

  Let us take a side step and ask out of curiosity the following question: 
Is it possible to construct a gauge-invariant Lagrangian with matter fields 
alone?  For instance, can we construct a {\em local} field theory with the 
electron-positron field alone such that it is invariant under the space-time 
dependent rotation $\psi(x) \to e^{i\alpha(x)}\psi(x)$ even in the absence 
of an auxiliary gauge field ?  If the particles are bosons, the 
$CP^N$/$CP^{N-1}$ model\cite{CPN} would probably be the best known example 
of this type. Its supersymmetric extension was also discussed.\cite{Akh}
In the case that the matter fields are fermions alone, the history actually 
goes much further back to the work by Bjorken\cite{Bj}, but the work along 
this line has not been fruitful.\footnote{A review is found for some of 
early history at the beginning of the Reference \cite{Akh} including 
references.}

  The method of the auxiliary vector fields was often used in the past to 
proceed in this kind of argument. It introduces nonpropagating gauge fields 
at start and their kinetic energy terms are added later by the loop 
contribution, ending up with the Lagrangian of matter \underline{and} 
propagating gauge fields.  Many argued that the nonpropagating gauge field 
implanted as an auxiliary field in Lagrangian should be interpreted as 
turning into a bound state once it has acquired its kinetic energy from 
the loop contributions.  But it is an inevitable consequence of gauge 
invariance of Lagrangian that such an auxiliary field, elementary or 
otherwise, ought to acquire a gauge invariant kinetic energy term 
$-\frac{1}{4}G_{\mu\nu}G^{\mu\nu}$ after loops are included.  Wouldn't it 
be more illuminating if composition of the massless vector-state can be seen 
explicitly in terms of the constituent matter fields ?  Such a diagrammatic 
computation was indeed made by Haber, Hinchliffe and Ravinovici\cite{HHR} 
for the $CP^{N-1}$ model many years ago. Unfortunately, this demonstration 
cannot be repeated when the constituents are fermions, since a simple 
{\em local} gauge-invariant Lagrangian corresponding to that of the 
$CP^N$ model has not been known in the case of fermion constituents. 

More recently, attempt has been made to introduce composite gauge bosons 
through the fifth dimension of the Randall-Sundrum model \cite{Raman}. 
The gauge bosons live in the branes and can be interpreted as composite 
wholly or partially. This is a new class or concept of composite gauge 
bosons.  Models were built and phenomenology was discussed for possible 
extensions of the standard model along this line.\cite{Rizzo, Gher}. 
    
In this paper we would like to focus on the dynamics of formation of 
composite gauge bosons at an elementary level of particle physics. 
Many of us have the underlying conviction or speculation that when 
Lagrangian is locally gauge invariant, gauge bosons must emerge as
composite states even if they are not placed as elementary particles. 
We would like to see it with our model Lagrangians in an explicit 
diagrammatic way.  In order to separate the issue from the argument 
based on the auxiliary vector field trick, we study the Lagrangians 
consisting of fermion fields alone. Furthermore, since our Lagrangian
consists only of fermions, supersymmetry is not relevant to our 
argument, barring the nonlinear realization\cite{Volk}. We stay in the 
flat space-time of dimension four all the time.  We have no need of an extra 
dimension explicitly or implicitly.  Given our Lagrangian, we can carry 
through diagram calculation in the leading 1/N order with no further 
approximation or assumption. In this way we can observe how the composite 
gauge bosons are made of their consituents dynamically. Our reasoning 
for construction of the Lagrangian is simple and resorts to no
sophisticated mathematical argument or technique. 

 The primary purpose of this paper is to give model Lagrangians that 
advocate inevitability of gauge bosons in gauge symmetric theories. 
Although application of our class of model Lagrangians to the real 
world is not our primary concern at this moment, short comments are 
made at the end on issues in electroweak phenomena. At the end, 
looking back the history of ``compositeness'' including findings
in some supersymmetric theories, we wonder if it is really a meaningful 
concept at a fundamental level.

At present, we do not have in mind an immediate application of our model  
Lagrangian to particle phenomenology.  The gauge bosons have been generally
accepted as the ``elementary'' particles and, experimentally, there is no 
compelling reason of compositeness for them.  Therefore we shall not pursue
experimental relevance of our models seriously in this paper.  Our emphasis
at present is primarily on their theoretical implications in composite
gauge bosons in general.  When Yang and Mills introduced the non-Abelian 
gauge field theory\cite{YM}, it had no immediate application. Even the 
$\rho$-meson was not known at that time although the concept of the weak
intermediate bosons was entertained by theorists. The Yang-Mills theory 
became a subject of intense 
phenomenological interest only after the Higgs mechanism\cite{Higgs}, 
Weinberg's ``A Model of Leptons''\cite{W} and quantum chromodynamics were 
developed unexpectedly one after another. If we recall this history,
we may have chance to see some feature of our models develop into a subject 
of experimental interest as the Large Hadron Collider upgrades luminosity 
and energy further.
 
We organize the paper as follows: In Section II, following the footstep of 
the $CP^N$ model, we introduce the U(1) gauge model of charged Dirac fields
alone. We emphasize that, in contrast to the $CP^N$ model, one cannot 
write a $\underline{local}$ Lagrangian of fermion fields alone with 
the so-called auxiliary field trick.    
In Section III we show that the Noether current is inevitably absent in 
the gauge theories that consist of matter fields alone.  In Section IV, 
we show dynamics of the U(1) gauge-boson formation first in the bosonic 
matter model and then in the fermionic matter model. We introduce, as usual, 
the N families of matter fields and take the large N limit in order to 
solve the models explicitly in a compact form.  We find that a massless 
bound state appears in the $^3S_1$ channel of elastic fermion-antifermion 
scattering, but that the fermion-antifermion pair must interact in the 
$^3D_1$ channel as well in order to form the massless bound state of 
spin-one.  In Section V we extend our models to the non-Abelian gauge 
symmetry. Choosing the matter fields in SU(2) doublet, we can build 
a non-Abelian model with Dirac fields. Computing the elastic scattering
amplitude, we find the non-Abelian gauge bosons in the SU(2)-triplet 
channel as bound states with the correct self-couplings as required by 
the non-Abelian gauge invariance.  In our class of models, the SU(2)-doublet 
matter plays a special role; it is impossible to extend the model to 
matter fields of general SU(2) multiplets nor to general Lie groups. 
The special role of the SU(2) doublet is discussed in the text and also 
with two examples in one of the Appendices. In the final Section VI, 
we discuss on relevance of the missing Noether currents to the no-go 
theorem of Weinberg and Witten\cite{WW}.  We conclude with comments
on possible relevance to the electroweak phenomenology and on historical
mutation of the concept of compositeness.

\section{U(1) Models}

  We proceed by following an elementary line of argument.  The first step 
is to construct a local Lagrangian $L(\psi, \overline{\psi})$ such that
\begin{equation}
 L(e^{i\alpha(x)}\psi(x), e^{-i\alpha(x)}\overline{\psi}(x)) = 
              L(\psi(x), \overline{\psi}(x)), \label{gaugeinv1}
\end{equation}
where $L(\psi(x),\overline{\psi}(x))$ depends on space-time coordinates 
$x_{\mu}$ only through the unconstrained fields $\psi(x)/\overline{\psi}(x)$. 
We cannot construct such a Lagrangian backward from the QED Lagrangian by 
integrating out the gauge field $A_{\mu}(x)$: We would need a gauge fixing 
to integrate over $A_{\mu}(x)$, but fixing a gauge breaks manifest gauge 
invariance.  We make our search here with the $CP^{N}$ model as a guide. 

  Quantum electrodynamics cannot be modified or extended in our way if both 
renormalizability and locality are required in the space-time of (3+1) 
dimensions. We do not consider here genuinely or intrinsically nonlocal 
field theories in which the fundamental fields and/or interaction contains 
nonlocality.\footnote{For example, the field theories once considered by 
Yukawa\cite{Y} and his followers.} In contrast to nonlocality, 
unrenormalizability can be controlled formally by dimensional regularization 
or by a covariant cutoff in phenomenology.  Therefore, we abandon here 
renormalizability in (3+1) dimensions for the moment and move to a world off 
(3+1) dimensions or consider a covariant cutoff theory in (3+1) dimensions. 
 
\subsection{Boson matter}

In order to construct a local Lagrangian with fermion matter fields 
alone, we first reexamine the gauge invariance of the bosonic matter
model, namely the $CP^N$ model, from a slightly different viewpoint.   
 
In the $CP^N$ model the gauge noninvariance of the free Lagrangian $L_0$ 
due to $\partial^{\mu}\phi$ under $\phi\rightarrow e^{i\alpha(x)}\phi$ 
must be counterbalanced with that of the interaction $L_{int}$. Therefore
$L_{int}$ must have at least the same number of derivatives as $L_0$. 
Since $L_0$ and $L_{int}$ have the same space-time dimension, we must 
introduce an inverse of $(\phi^*\phi)$ in $L_{int}$ to make up for the 
dimension due to $\partial^{\mu}$ in the numerator of $L_{int}$. 
Keeping the number of $\partial^{\mu}$ in $L_{int}$ the smallest, we 
reach almost uniquely the simplest form of the gauge-invariant
 Lagrangian made of the matter fields alone as
\begin{equation}
    L_{tot} = L_0 + L_{int}, \label{Lb}
\end{equation}
where $L_0$ is the standard free Lagrangian,
\begin{equation}
    L_0 =  \sum_{i=1}^N\partial^{\mu}\phi_i^*\partial_{\mu}\phi_i
                 - \sum_{i=1}^N m^2\phi_i^*\phi_i,  \label{Lb0}
\end{equation}
and the interaction Lagrangian $L_{int}$ is given by 
\begin{equation}
    L_{int}  = \lambda  
   \frac{\sum_{i=1}^N(\phi_i^*\stackrel{\leftrightarrow}{\partial}^{\mu}\phi_i)
     \sum_{j=1}^N(\phi_j^*\stackrel{\leftrightarrow}{\partial}_{\mu}\phi_j)}
    {4\sum_{k=1}^N(\phi^*_k\phi_k)}, \;\; (\lambda\rightarrow 1).     \label{Lbint}
\end{equation}
The indices $(i,j,k)$ run from $1$ to $N$ so that the model be solvable 
in the leading order of $1/N$.  They are referred to as the {\em copy} 
indices hereafter.  From time to time, however, the summation over the
copy indices will be suppressed unless we need to remind of it. 

Under the local U(1) gauge transformation, the fields transform with 
a space-time dependent phase $\alpha(x)$ common to all the copy index $i$ as
\begin{equation}
    \phi_i \to e^{i\alpha(x)}\phi_i, \;\; \mbox{and} \;\; 
    \phi_i^* \to e^{-i\alpha(x)}\phi_i^*.               \label{bgt}
\end{equation}
For the total Lagrangian, each of $L_0$ and $L_{int}$ varies nontrivially under 
the gauge transformation Eq.(\ref{bgt}), but the variations $\delta L_0$ and 
$\delta L_{int}$ are so made as to be proportional to each other:
\begin{eqnarray}
 \delta L_0 &=& -i\Bigl(\sum_i\phi_i^*\stackrel{\leftrightarrow}{\partial}_{\mu}\phi_i\Bigr)
   \partial^{\mu}\alpha +\Bigl(\sum_i\phi_i^*\phi_i\Bigr)
                          \partial^{\mu}\alpha\partial_{\mu}\alpha,   \nonumber \\
       \delta L_{int} &=& - \lambda\delta L_0.  \label{cancel}
\end{eqnarray} 
These gauge variations cancel each other between $L_0$ and $L_{int}$ for 
\begin{equation}
               \lambda =1 \;\;(\mbox{gauge limit}).       \label{gaugelimit}
\end{equation}
If we remove the mass term and impose the constraint $\sum_i\phi^*_i\phi_i = N/2f$ 
in Eq.(\ref{Lbint}), we recognize this Lagrangian (with $\lambda = 1$) as that of 
the $CP^{N-1}$ model \cite{CPN}. However, we have introduced N copies solely for
the purpose of the computational ease of the leading 1/N expansion. Our interest 
is not in the SU(N) symmetry among the different copies.

As far as U(1) gauge invariance is concerned, we may add to Eq.(\ref{Lb}) the 
terms that are gauge invariant by themselves.  For instance, nonderivative 
$\phi^4$-couplings such as 
\begin{equation}
   L'_{int} = - \sum_{i,j=1}^N\lambda_{ij}(\phi_i^*\phi_i)(\phi_j^*\phi_j), 
                            \label{phi-fourth}
\end{equation}
where $\lambda_{ij}$ are arbitrary real constants. However, in the leading 1/N 
order the interactions such as $L'_{int}$ do not affect on the bound-state 
formation.\footnote{Because we compute the bound state of spin-one, not of 
spin-zero.}  Therefore we leave out such interactions hereafter.  It is 
reassuring to see later that the vector bound state comes out massless with 
the correct gauge coupling irrespectively of the additional gauge-invariant
interactions such as $L_{int}'$.

\subsection{Fermionic model}
 
   Following the reasoning outlined above, we can obtain with a little 
stretch of imagination a fermionic extension of the bosonic model Lagrangian 
Eq.(\ref{Lb}).  Since the free Lagrangian 
$L_0$ contains only one first-derivative of $\psi$, the interaction $L_{int}$ 
can counterbalance the gauge variation of $L_0$ with only one first-derivative 
of field.  Just as in the bosonic case, we need to introduce the inverse of 
the scalar density $\overline{\psi}\psi$ in $L_{int}$ in order to match the 
dimension.  Following this reasoning as in the bosonic model, we reach 
the Lagrangian $L_0 + L_{int}$,
\begin{eqnarray}
  L_0 &=&\sum_i\overline{\psi}_i(i\!\not\partial - m)\psi_i, \nonumber \\ 
  L_{int}&=& -i\lambda\frac{\sum_i(\overline{\psi}_i\gamma_{\mu}\psi_i)        
     \sum_j(\overline{\psi}_j\stackrel{\leftrightarrow}{\partial}^{\mu}\psi_j)}{ 
   2\sum_k\overline{\psi}_k\psi_k},  \;\; (\lambda\rightarrow 1),  \label{Lf}  
\end{eqnarray}
where the gauge invariance is realized at $\lambda = 1$. 
 Under the gauge transformation,
\begin{eqnarray}
     \psi &\to& e ^{i\alpha(x)}\psi   \nonumber \\
      \overline{\psi} &\to& \overline{\psi} e^{-i\alpha(x)} , \label{fgt}
\end{eqnarray}
the Lagrangian of Eq.(\ref{Lf}) is invariant by cancellation between
the gauge variations of $L_0$ and $L_{int}$ at $\lambda = 1$:  
\begin{eqnarray}
     \delta L_0 &=& -\overline{\psi}(\not\!\partial\alpha)\psi, \nonumber \\
 \delta L_{int}&=&\lambda\overline{\psi}(\not\!\partial\alpha)\psi.\label{fgtL}
\end{eqnarray}
We may add to $L_{int}$ the self-gauge-invariant terms such as
\begin{equation}
  L'_{int} =
 -\frac{fm}{4}(\overline{\psi}\gamma_{\mu}\psi)\frac{1}{(\overline{\psi}\psi)}
             (\overline{\psi}\gamma^{\mu}\psi),     \label{Lfprime}  
\end{equation}
where insertion of the fermion mass $m$ is just to make the constant $f$ 
dimensionless. The constant $f$ is unconstrained by gauge invariance.  
After we compute for the massless bound state with $L_{int}$ of Eq.(\ref{Lf}) 
alone, we shall examine how the interactions like $L'_{int}$ affect its mass 
and coupling. Since they will turn out to be irrelevant to determination of 
the mass and coupling of the massless bound state, we shall not include 
them in our diagram calculation.  Before diagram calculation, some may 
suspect that the fermion-antifermion interaction through 
$\propto (\overline{\psi}\gamma_{\mu}\psi) (\overline{\psi}\gamma^{\mu}\psi)$ 
might be responsible for or relevant to binding a gauge boson. It is wrong. 
Such an interaction does not exist in our $L_{int}$.  Even if one includes it 
in $L_{int}$, it does not participate in formation of the massless gauge boson 
nor in determination of the gauge coupling, as we shall see later.

Our fermionic Lagrangian Eq.(\ref{Lf}) is obviously nonrenormalizable 
in four space-time dimensions just like that of the $CP^{N}$ model. 
As we know, the only renormalizable U(1) gauge field theory with a charged 
fermion is quantum electrodynamics: It needs the propagating gauge field 
$A_{\mu}$ explicitly in Lagrangian. 

\subsection{Auxiliary vector-field trick}

Our bosonic Lagrangian Eq.(\ref{Lb}) with $\lambda=1$ takes the 
same form as what we could obtain by starting with the gauge-invariant 
Lagrangian of a nonpropagating auxiliary gauge field $A_{\mu}$,
\begin{equation}
 L_{aux} = \sum_i(\partial_{\mu}-ieA_{\mu})\phi_i^*(\partial^{\mu}+ieA^{\mu})\phi_i
                                  -m^2\phi_i^*\phi_i.   \label{auxb}
\end{equation}
Either by integrating Eq.(\ref{auxb}) over $A^{\mu}$ or by substituting the
equation of motion for $A^{\mu}$,
\begin{equation}
  eA_{\mu} = \frac{i}{2}(\sum_i\phi_i^*\stackrel{\leftrightarrow}
              {\partial}_{\mu}\phi_i)/(\sum_j\phi_j^*\phi_j),   \label{EofM}
\end{equation}
we obtain for $m^2\to 0$ the $CP^N$ Lagrangian (before imposing the 
constraint and turning it into $CP^{N-1}$)\cite{BKY}. 

  When we compute by the loop correction the dimension-four operator of 
$A^{\mu}$  for the effective action, we obtain the ``kinetic energy term'' 
$-\frac{1}{4}F_{\mu\nu}F^{\mu\nu}$.  One cannot obtain anything other than 
the gauge invariant $FF$ term (``the Maxwell term'') since the Lagrangian 
Eq.(\ref{auxb}) is gauge invariant by construction. Whether this appearance 
of the $FF$ term is to be interpreted as ``generation of a bound state'' 
or not should be subject to debate. If we accepted such interpretation, 
a massless spin-one state would emerge irrespectively of strength of the 
interaction $e^2$ which is implanted in Eq.(\ref{auxb}) at the beginning.  
After a rescaling of the $A_{\mu}$ field, the physical coupling of $A_{\mu}$ 
to $\phi/\phi^*$ is fixed to some number, which is independent of $e$ in 
the one-loop and logarithmically divergent in four dimensions. 
Turning of the field $A_{\mu}$ into a massless boson is guaranteed
once the field is introduced as an auxiliary field. In contrast, in our 
model the strength of interaction $L_{int}$ must be tuned to the optimum 
value ($\lambda = 1$) in order to make the bound state massless. In this
way we see that masslessness of the vector bound-state is a dynamical 
consequence of gauge invariance rather than a kinematical outcome.

Substitution of the equation of motion Eq.(\ref{EofM}) also needs scrutiny:
If one computes $\partial_{\mu}F^{\mu\nu}$ with this $A^{\mu}$, one would 
obtain $\partial_{\mu}F^{\mu\nu}= 0$ instead of 
$\partial_{\mu}F^{\mu\nu} = J^{\nu}$.  Therefore, the field $A^{\mu}$ of 
Eq.(\ref{EofM}) is not acceptable as the composite gauge field.  One 
would need contributions from loops to write a dynamical gauge field 
that obeys the correct equation of motion. We do not know how to write 
such an object in a local composite field.  

What would happen if one attempts to introduce the auxiliary field 
$A_{\mu}$ in the fermionic model ?  For the fermionic matter, the 
Lagrangian with a nonpropagating auxiliary field is simply equal to 
\begin{equation}
  L_{aux} = \sum_i\overline{\psi}_i(i\!\not\partial +e\!\not\!A - m)\psi_i,
                                             \label{auxf}
\end{equation}
The equation of motion with respect to $A_{\mu}$ is trivially equal to
$\sum_i\overline{\psi}_i\gamma_{\mu}\psi_i=0$ and provides us nothing.  
As for the functional integration over the auxiliary field $A_{\mu}$, 
one cannot carry it out at the tree level since the auxiliary Lagrangian 
Eq.(\ref{auxf}) is not quadratic in $A_{\mu}$ unlike that of the bosonic 
model.  When the two-point loop-diagrams of $A_{\mu}A_{\nu}$ is computed, 
the local limit of the two-point functions ought to be proportional 
to $F_{\mu\nu}F^{\mu\nu}$ by the underlying gauge invariance.  But we cannot
obtain a compact local Lagrangian of the matter fields alone such 
as ours out of the auxiliary Lagrangian of Eq.(\ref{auxf}). 

The auxiliary vector-field trick bypasses the important part of
dynamics of the matter fields. In contrast, our explicit Lagrangian
models provide dynamical details of binding which are missing in the 
auxiliary field trick or else very different from it.  

\section{Noether current}
When we attempt to write a conserved current in our models, we encounter 
one peculiar problem: We are unable to construct a conserved current with 
the prescription of the Noether theorem. In fact, such a current simply 
does not exist.

According to the general prescription, the Noether current $J_{\mu}^N$ is 
obtained when Lagrangian is invariant under a set of space-time independent 
phase transformations of fields. In the bosonic model, it would be generated 
by the transformation,
\begin{equation}
  \phi_i\to (1+i\alpha)\phi_i \;\; {\rm and} 
      \;\;\phi_i^*\to (1-i\alpha)\phi_i^*, \label{gtc} 
\end{equation}
where $\alpha$ is infinitesimal and {\em independent} of space-time.
The variation $\delta L_{tot}$ of $O(\alpha)$ under this transformation 
leads to divergence of the Noether current through the identification,
\begin{equation}
   \partial^{\mu} J_{\mu}^N = -\delta L_{tot}/\delta\alpha. \label{Noether}
\end{equation}
Using the equation of motion in the right-hand side, one ought to obtain
the Noether current $J_{\mu}^N$ as         
\begin{equation}
    J_{\mu}^N = 
      -i\sum_i\Bigl(\frac{\partial L_{tot}}{\partial(\partial^{\mu}\phi_i)}\phi_i 
      -\phi_i^*\frac{\partial L_{tot}}{\partial(\partial^{\mu}\phi_i^*)}\Bigr).
                                                  \label{NoetherEq}
\end{equation} 
When we follow this standard procedure in our models, we find that the 
right-hand side of Eq.(\ref{NoetherEq}) is identically zero in the gauge 
symmetry limit by cancellation between the contributions from $L_0$ 
and $L_{int}$:
\begin{equation}
     J_{\mu}^N  = i(1-\lambda)\sum_i(\phi_i^*\stackrel{\leftrightarrow}{\partial}_{\mu}
                                \phi_i),  \label{conservedb}
\end{equation}
where the term proportional to $\lambda$ comes from $L_{int}$ and the gauge 
symmetry holds at $\lambda = 1$.  
One may be puzzled when one thinks of perturbative calculation: Since $\phi$ 
and $\phi^*$ always appear pairwise in product in the Lagrangian, one may 
assign the conserved U(1) charge $\pm 1$ to $\phi$ and $\phi^*$.  Then this 
charge ought to be conserved in all diagrams of physical processes such as 
scattering and decay even in the gauge symmetry limit where the Noether current 
disappears. 

The same happens in the fermionic model too. Just as in the bosonic model, 
the conserved Noether current disappears in the gauge symmetry limit: 
\begin{equation}
     J_{\mu}^N  = (1-\lambda)\sum_i\overline{\psi}_i\gamma_{\mu}\psi_i.  
                   \label{conservedf}
\end{equation}
The current $\sum\overline{\psi}_i\gamma_{\mu}\psi_i$ is not the Noether 
current. It is a general property of the gauge theories having no gauge 
field that the Noether current is identically zero; $J_{\mu}^N\equiv 0$. 
It is easy to trace the root cause of absence of the Noether current 
to local gauge invariance itself.  An almost trivial proof is given in the 
Appendix A. The proof can be easily extended to the non-Abelian models.
It has an important implication in the non-Abelian case: If the Noether 
current existed, generation of the massless gauge bosons would face 
a potential conflict with the no-go theorem of Weinberg and Witten\cite{WW}.

   Unlike the Noether current, the conserved energy-momentum tensor exists 
in the Abelian and non-Abelian gauge theories of matter fields alone.
For the fermionic U(1) model with the Lagrangian of Eq.(\ref{Lf}), the 
conserved energy-momentum tensor is given by
\begin{equation}
 T^{\mu\nu}= i\sum_i\overline{\psi}_i\gamma_{\mu}\partial_{\nu}\psi_i -
   \frac{i\lambda(\sum_i\overline{\psi}_i\gamma^{\mu}\psi_i)
   (\sum_j\overline{\psi}_j\stackrel{\leftrightarrow}{\partial}^{\nu}\psi_j)}{
   2\sum_k(\overline{\psi}_k\psi_k)} -g^{\mu\nu}L_{tot}.  \label{Tf}
\end{equation}
It is manifestly gauge invariant with the matter fields alone. 

\section{Composite U(1) gauge boson}

It is natural to wonder if our U(1) models contain a gauge boson as a 
composite state even though we have not placed it by hand. In order to 
answer to this question, we carry out diagram calculation in this section 
in order to exhibit the dynamical mechanism of forming the composite 
gauge boson. We compute our models perturbatively in the 1/N expansion: 
We sum an infinite series of the leading 1/N order terms and show 
explicitly that a massless vector boson indeed appears as a pole in 
scattering amplitudes with the properties required by gauge symmetry 
both in the bosonic and the fermionic model. In the case of the 
$CP^{N-1}$ model in which $\phi^*\phi$ is subject to a constraint, 
this diagram computation was done by Haber {\em et al} \cite{HHR}. 
Our primary interest is in the fermionic model, which is technically 
complex since channel coupling occurs between the $^3S_1$ and $^3D_1$ 
channels. Unlike the formal argument based on the auxiliary 
vector-field trick\cite{RS}, the diagrammatic computation allows 
us to see explicitly how a massless bound state is formed dynamically 
with the matter particles.  For instance, when we examine elastic 
fermion-antifermion scattering of $J^{PC}=1^{--}$, we find that 
the massless bound state appears in the $^3S_1$ channel, not in the 
$^3D_1$ channel. That is, the bound state couples with the fermions 
through the vertex $\overline{\psi}\gamma_{\mu}\psi$, not through 
$\overline{\psi}\stackrel{\leftrightarrow}{\partial}_{\mu}\psi$.
Nonetheless, the interactions of both types are needed to form a 
massless bound state.

\subsection{Gauge boson in bosonic model}

   We start with our U(1) bosonic model for study of a composite 
gauge boson before our study of the fermionic model since the 
computation is simpler for the bosonic model, yet it demonstrates 
essential aspects of the diagram calculation.

   We consider the two-body $\phi^+\phi^-$ scattering in $p$-wave 
($J^{PC}=1^{--}$), treating all $N$ copies of the fields ($i=1,\cdots N$) 
as independent. We show that a pole of a massless bound-state appears 
in this channel.  Then we proceed to make sure that the pattern 
and magnitude of the coupling of this bound state indeed obey what 
we expect for the U(1) gauge boson.  

   We study the $p$-wave amplitude for the two-body scattering, 
\begin{equation}
    \phi_i^+(p_1)+\phi_i^-(p_2) \rightarrow \phi_j^+(p_3) + \phi_j^-(p_4).
                                   \label{Bscat}
\end{equation}
We compute the amplitude in the leading $1/N$ order since a compact explicit 
solution can be obtained only in this order.  In the scattering Eq.(\ref{Bscat}), 
the copy indices are chosen to be the same for the initial particles and also 
for the final particles. In the diagram calculation, $L_{int}$ is separated 
from $L_{tot}$ in Eq.(\ref{Lbint}) and treated as {\em the interaction}.  While 
this statement sounds trivial, we point out one subtlety. That is, when we carry 
out perturbative calculation by splitting the Lagrangian into $L_0$ and $L_{int}$, 
we have fixed once for all the gauge ambiguity of our Lagrangian Eq.(\ref{Lb}).   
That is, when we write the propagator of $\phi/\phi^*$ in the momentum space as 
$1/(p^2-m^2)$, we need no more gauge fixing since there is no $A^{\mu}$ field 
in the Lagrangian.  With this separation, the fields obey the equation of 
motion of $L_0$ that violates gauge symmetry. Consequently the Noether current
of $L_0$ is the conserved current in diagrams.  For the purpose of visualizing 
how the gauge-invariance limit is reached, we float $\lambda$ in $L_{int}$ as 
a free parameter until we set it to unity at the end of calculation.

In the diagram calculation of the leading 1/N order, we normal-order the 
operator $\phi^*\phi$ in the denominator of $L_{int}$ and expand it around 
its vacuum value as
\begin{eqnarray}
  1/\sum\phi^*\phi &=& 
                1/\Bigl(\sum \langle 0|\phi^*\phi|0\rangle + \sum:\phi^*\phi:\Bigr) 
                                                     \nonumber \\
                 &=& \frac{1}{\sum\langle 0|\phi^*\phi|0\rangle}
                     \sum_{n=0}^{\infty}(-1)^n\Bigl(\frac{\sum:\phi^*\phi:}{
                         \sum\langle 0|\phi^*\phi|0\rangle}\Bigr)^n,
                                                    \label{expansion}
\end{eqnarray}
where the summation $\sum$ with no index attached means the summation over the copy 
index $i(=1,\cdots N)$. This separation of the vacuum value is important to handle
the denominator of $L_{int}$ in a systematic 1/N expansion.\cite{HHR}
The vacuum expectation value $\langle 0|\phi^*\phi|0\rangle$ 
is infinite in the (3+1) space-time, so it is regularized dimensionally as
\begin{eqnarray}
  \sum\langle 0|\phi^*\phi|0\rangle &=& \lim_{x\rightarrow 0} 
              \sum \langle 0|T(\phi^*(x)\phi(0))|0\rangle, 
                                       \nonumber \\
       &=& \frac{N\Gamma(1-D/2)}{(4\pi)^{D/2}(m^2)^{1-D/2}}, \label{vacb}
\end{eqnarray}
where $N$ copies of bosons contribute to the vacuum value of the scalar 
density. The space-time dimension $D$ is set to four eventually.
We denote this vacuum-expectation-value by $I^b_0$ hereafter,
\begin{equation}
     I_0^b \equiv \sum \langle 0|\phi^*\phi|0\rangle.             \label{vac1}
\end{equation}
Now we are ready to compute for the two-body scattering of Eq.(\ref{Bscat}).
The great simplification of the leading $1/N$ order is that for elastic 
scattering we have only to sum the chain of the bubble diagrams, as shown 
in Fig. 1, in which the copy index $i$ runs within a loop of each bubble. 

\noindent
\begin{figure}[h]
\epsfig{file=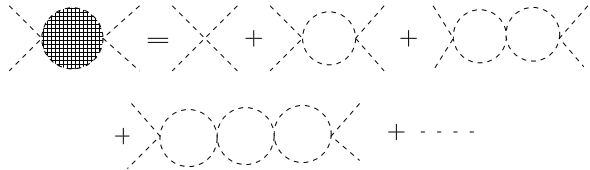,width=0.47\textwidth}
\caption{The chain of the bubble diagrams for the elastic boson scattering.
\label{fig:1}}
\end{figure}
 
Let us define with the S-matrix the two-body scattering amplitude 
$T(p_3,p_4;p_1,p_2)$ as
\begin{equation}
     <p_3,p_4|S-1|p_1,p_2> =i(2\pi)^4 \delta^4(p_3+p_4-p_1-p_2)
                     T(p_3,p_4;p_1,p_2).  \label{S}
\end{equation}
The amplitude $T$ has the Lorentz structure of the form
\begin{equation}
    T(p_4,p_3;p_1,p_2) = (p_3-p_4)^{\mu}(p_1-p_2)^{\nu}T(q)_{\mu\nu}, \label{T}
\end{equation}
where $q=p_1+p_2 = p_3+p_4$ and the one-particle states are normalized as 
$\langle p_i|p_j\rangle =2E_i\delta(\mbox{\bf p}_i- \mbox{\bf p}_j)$ so that 
the amplitude $T(p_3,p_4;p_2,p_1)$ is a Lorentz scalar. For the elastic 
scattering in the leading $1/N$ order, it is sufficient to keep only the first 
term of the expansion Eq.(\ref{expansion}) in the denominator of $L_{int}$. 
The normal-ordered product $(\sum :\phi^*\phi:)$ starts contributing to the 
next-to-leading order of 1/N in the elastic scattering.

     The amplitude $T(q)_{\mu\nu}$ starts with a contact interaction term with 
no bubble, the first term in the right-hand side of Fig. 1, which is equal to
\begin{equation}
    T^0_{\mu\nu} = \frac{\lambda}{2I^b_0}g_{\mu\nu},  \label{Bornb}
\end{equation} 
where the superscript ``zero'' of $T^0_{\mu\nu}$ indicates the zero-loop 
contribution of $O(\lambda)$. The bubble summation can be carried out by 
solving the algebraic equation (Fig. 2),
\begin{equation}
 T(q)_{\mu\nu} =T^0_{\mu\nu} + K(q)_{\mu\kappa}T^{\kappa}_{\nu}(q). 
                                        \label{bAlEq}
\end{equation}
where the kernel $K(q)_{\mu\kappa}$ is given by the single bubble diagram 
in which the copy index flows around the loop.  Eq.(\ref{bAlEq}) will 
become powerful later when we sum the corresponding series in the fermionic 
model in which two eigenchannels contribute and entangle in formation of 
a bound state. 

\noindent
\begin{figure}[h]
\epsfig{file=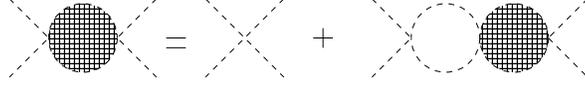,width=0.47\textwidth}
\caption{The iteration equation of bubbles into a chain. 
\label{fig:2}}
\end{figure}
 Straightforward computation gives us the kernel as 
\begin{equation}
 K_{\mu\kappa}(q) =
 \frac{\lambda N \Gamma(1-D/2)}{(4\pi)^{D/2}(m^2)^{1-D/2}I^b_0}\Bigl(g_{\mu\kappa} 
  +\frac{1-D/2}{6m^2}(g_{\mu\kappa}q^2 -q_{\mu}q_{\kappa})\Bigr) +O(q^4). 
                             \label{kernel}  
\end{equation}
Since we want to extract the pole and residue of a massless bound state
at $q^2 = 0$, we need $K_{\mu\kappa}(q)$ only to the orders no higher 
than $O(q^2)$. 
The factor outside the large bracket in Eq.(\ref{kernel}) is simply equal to 
$\lambda$ when Eq.(\ref{vacb}) is substituted for $I_0^b$ so that
\begin{equation}
K_{\mu\kappa}(q) = \lambda \Bigl(g_{\mu\kappa}
    +\frac{1-D/2}{6m^2}(g_{\mu\kappa}q^2 -q_{\mu}q_{\kappa})\Bigr)
    +O(q^4). \label{kernel1}  
\end{equation}
Note here that $K_{\mu\kappa}(q)$ does not satisfy the transversality, 
$q^{\mu}K_{\mu\kappa}\neq 0$.  This is not violation of gauge invariance.  
In the standard Lagrangian where the elementary $A_{\mu}$ field is present, 
one would need the $A_{\mu}A^{\mu}\phi^*\phi$ term to realize transversality 
of the photon self-energy, $q^{\mu}\Pi(q)_{\mu\kappa}=0$, namely, gauge 
invariance. The term needed for transversality does exist in our model, but 
it is tucked away elsewhere at this stage.  As we shall see in a moment, 
it is this nontransversality of $K_{\mu\kappa}(q)$ that makes the composite 
boson massless.\footnote{This is the case in the $CP^{N-1}$ model 
analyzed in \cite{HHR} too.}.  

Let us substitute Eq.(\ref{kernel1}) in the iteration equation 
Eq.(\ref{bAlEq}) and move the term $\lambda g_{\mu\kappa}$ of 
the kernel $K_{\mu\kappa}(q)$ to the left-hand side.  We may drop 
the term proportional to $q_{\mu}q_{\mu}$ by use of $q\cdot(p_1-p_2)=0= 
q\cdot(p_3-p_4)$ on the external boson lines.  Then Eq.(\ref{bAlEq}) 
turns into
\begin{equation}
 (1-\lambda)T(q)_{\mu\nu} =T^0_{\mu\nu} + 
         \frac{\lambda(1-D/2)q^2}{6m^2} T(q)_{\mu\nu} +O(q^4). 
                                        \label{AlEq1}
\end{equation}
Now we go to the gauge limit $\lambda\rightarrow 1$. Since $T_{\mu\nu}^0$ is
independent of $q$, Eq.(\ref{AlEq1}) tells us that in this limit there is 
a pole at $q^2=0$ in the amplitude $T(q)_{\mu\nu}$ as
\begin{equation}
          T(q)_{\mu\nu} = -\frac{6m^2}{(1-D/2)q^2}T^0_{\mu\nu} +O(q^2),
                        \;\;\;(\lambda = 1).     \label{pole}
\end{equation}
When the parameter $\lambda$ is off the gauge limit ($\lambda\neq 1$), 
the pole is located away from zero at $q^2= [6(1-\lambda)/\lambda(1-D/2)]m^2$ 
so that the bound state would be either a massive vector boson or a tachyon.  
We extract the residue of the pole at $q^2 = 0$ for $\lambda = 1$ and compare 
this residue with what we would obtain from the Feynman diagram of the standard 
U(1) gauge Lagrangian of the charged scalar fields, 
\begin{equation}
     L_{tot} = -\frac{1}{4}F_{\mu\nu}F^{\mu\nu} +
(\partial^{\mu}\phi^*-ieA^{\mu}\phi^*)(\partial_{\mu}\phi +ieA_{\mu}\phi) 
                - m^2\phi^*\phi.     \label{U1bLagrangian}
\end{equation} 
By equating our residue with that of Feynman diagram, we obtain the coupling $e^2$ 
of our model as
\begin{equation}
        e^2 = \frac{3(4\pi)^{D/2}(m^2)^{2-D/2}}{N\Gamma(2-D/2)}. \label{U1bcoupling}
\end{equation}
When we approach the space-time dimension of $D=4$, this coupling can be 
expressed in terms of the logarithmic cutoff of divergence as
\begin{equation}
        e^2 = \frac{48\pi^2}{N\ln({\overline{\Lambda}^2/m^2)}}, \label{U1bcoupling1}
\end{equation} 
where $\ln\overline{\Lambda^2} = (2-D/2)^{-1}+\ln 4\pi -\gamma_E$ ($\gamma_E$ = 
Euler constant). The sign of $e^2$ comes out to be correctly positive. 
It is amusing to observe that the factor $(1-D/2)$ in the denominator of
Eq.(\ref{pole}) is combined with $\Gamma(1-D/2)$ in $1/I_b^0$ of $T_{\mu\nu}^0$
to turn into $\Gamma(2-D/2)$, which is the logarithmic divergence in the
space-time dimension of $D=4$.  That is, a quadratic divergence $\Gamma(1-D/2)$ 
metamorphoses into a logarithmic divergence, which can happen in the dimensional
regularization. 

If we started with the auxiliary $A_{\mu}$ field and generate the $-\frac{1}{4}
F_{\mu\nu}F^{\mu\nu}$ to the leading 1/N order, we would obtain the coupling 
constant identical with Eq.(\ref{U1bcoupling1}) after rescaling $A_{\mu}$ by 
wave-function renormalization.\cite{Akh}  This equality is not unexpected since 
the one-loop self-energy diagram of the auxiliary $A_{\mu}$ field leading to 
Eq.(\ref{U1bcoupling1}) is identical with the bubble diagram of the $p$-wave 
$\phi^{\dagger}\phi$ scattering in the leading 1/N order. There is no guarantee 
that this equality holds beyond the leading 1/N order since noncontact 
interactions enter the scattering amplitude while the self-energy diagram 
remains the two-point function.

  In order to claim that the massless bound state discovered above is 
indeed the U(1) gauge boson, we must show that other couplings 
of this state obey the pattern required for the gauge boson. One may 
bypass this part by resorting to the gauge invariance that has been 
embedded in the Lagrangian of our model.  But we show here explicitly
how the U(1)-gauge invariance arises diagrammatically for the coupling
of the massless bound state.  
 
Absence of the coupling $eA_{\mu}\partial^{\mu}(\phi^*\phi)$ is obvious since 
the form of our $L_{int}$ requires the bound state to couple with $\phi^*/\phi$ 
through $(\phi^*\stackrel{\leftrightarrow}{\partial}^{\mu}\phi)$ not through
$\partial^{\mu}(\phi^*\phi)$. This is also required by $C$-invariance of 
our Lagrangian.  However, there must exist the coupling $e^2\phi^*\phi 
A_{\mu}A^{\mu}$, where $A_{\mu}$ is the effective gauge field and $e^2$ is 
to be given by Eq.(\ref{U1bcoupling}). Aside from this coupling, there 
should be no coupling of dimension four such as a nonderivative quartic 
coupling of $A_{\mu}$. 

The coupling of $\phi^*\phi A_{\mu}A^{\mu}$ requires a little computation. 
Here the first nontrivial term of the expansion of $1/(\phi^*\phi)$ 
enters the computation,
\begin{equation}
   -\frac{\lambda}{4I_0^2}(\phi^*\stackrel{\leftrightarrow}{\partial}^{\mu}\phi)
     (\phi^*\stackrel{\leftrightarrow}{\partial}_{\mu}\phi)(:\phi^*\phi:).
                                           \label{sixbody}
\end{equation}
In the leading 1/N order, we attach a chain of the bubble diagrams to 
$(\phi^*\stackrel{\leftrightarrow}{\partial}^{\mu}\phi)$ and another chain
to $(\phi^*\stackrel{\leftrightarrow}{\partial}_{\mu}\phi)$ to form 
the composite $A^{\mu}$ and $A_{\mu}$ bosons, respectively. (See Fig. 3.)  
Then we equate this diagram at the poles of the $A^{\mu}$ and $A_{\mu}$ 
bosons to the diagram of Fig. 4 which is obtained with the interaction 
$e^2\phi^*\phi A_{\mu}A^{\mu}$ of the standard U(1) gauge Lagrangian, 
Eq.(\ref{U1bLagrangian}). 

\noindent
\begin{figure}[h]
\epsfig{file=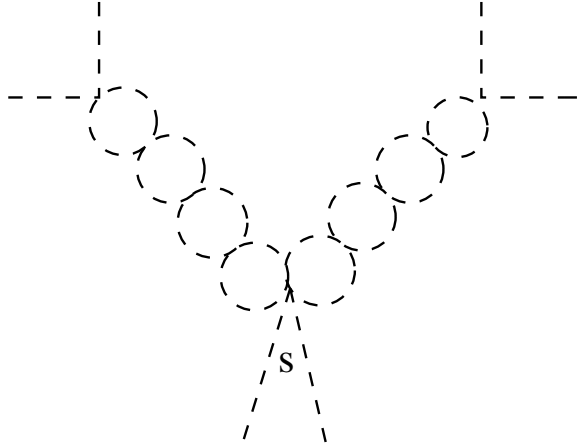,width=0.47\textwidth}
\caption{The diagram for formation of $\phi^*\phi A_{\mu}A^{\mu}$ coupling.
The $\phi^*\phi$ pair arises from the six-body interaction of 
Eq.(\ref{sixbody}) at the center.
The letter $S$ denotes that the external $\phi^*\phi$ pair at the center 
is in the scalar state $\phi^*\phi$, not in the vector state 
$\phi^*\stackrel{\leftrightarrow}{\partial}_{\mu}\phi$.\label{fig:3}} 
\end{figure}

\noindent
\begin{figure}[ht]
\epsfig{file=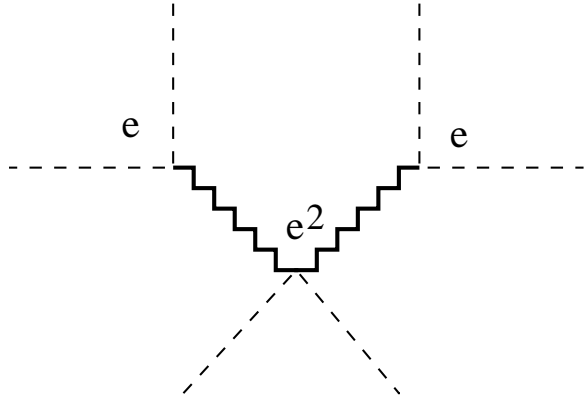,width=0.47\textwidth}
\caption{The corresponding Feynman diagram for $e^2\phi^*\phi A_{\mu}A^{\mu}$.
\label{fig:4}}
\end{figure}

This calculation gives us the relation 
\begin{equation}
    e^4 = \Bigl(\frac{3(4\pi)^{D/2}(m^2)^{2-D/2}}{
        N\Gamma(2-D/2)}\Bigr)^2.\label{AAphiphi}
\end{equation}
Two powers $e^2$ out of $e^4$ in Eq.(\ref{AAphiphi}) are to be attributed to 
the couplings of the $\phi^*\phi$ pairs with $A_{\mu}$ and with $A_{\nu}$ at the 
outer ends of two bubble chains in Fig. 3.  The remaining $e^2$ is to be assigned 
to the four-body $A_{\mu}A^{\mu}\phi^*\phi$ coupling at the center. Therefore, 
the coupling $e^4$ of Eq.(\ref{AAphiphi}) is precisely what we want to see.

  Absence of the triple self-coupling of $A_{\mu}$ is a consequence 
of $C$-invariance.  Diagrammatically, this is assured in the U(1) model 
by cancellation between a pair of diagrams where the two chains are
interchanged.  Since they do not cancel in the non-Abelian models and 
there is some subtlety, we add a few comments here in anticipation of 
the non-Abelian cases. The relevant diagram is depicted in Fig. 5. 

\begin{figure}[ht]
\epsfig{file=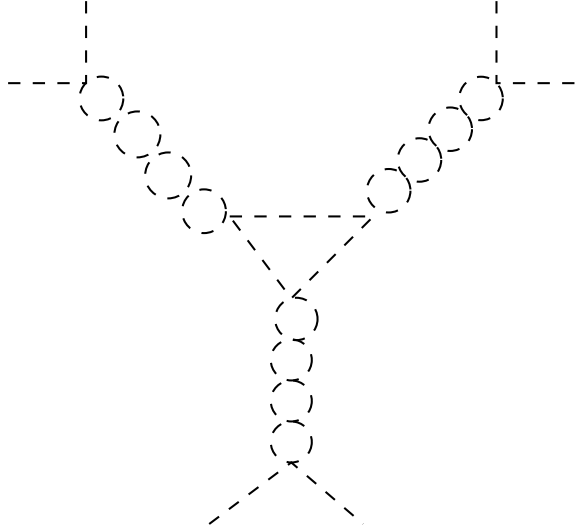,width=0.47\textwidth}
\caption{The triple self-coupling of the composite $A_{\mu}$, which can 
appear potentially from the center of the diagrams containing three 
chains of $\phi^*/\phi$ bubbles.
\label{fig:5}}
\end{figure}

 If we indeed compute this coupling with individual diagrams, we
must be careful about the surface-term ambiguity.  The triangular loop 
at the center is linearly divergent in the space-time dimension of four 
and therefore its constant term is ambiguous by the surface term 
of loop-integral. The value depends on how the loop-momentum is routed 
just as in the chiral anomaly or the finite part of the electron 
self-energy in QED. To fix this finite ambiguity, one must impose 
invariance and/or symmetry that must be preserved in theory.  In this 
case $C$ invariance of $L_{tot}$ and/or the Bose statistics of 
the composite $A_{\mu}$ fixes the ambiguity.  With the right choice of 
the routing momentum, a pair of triangular loop diagrams cancel each 
other and turn the net triple self-coupling to zero in the U(1) model. 

In comparison, we need an explicit computation of diagrams to 
show that the net quartic self-coupling vanishes, although there 
is no subtlety of the surface-term ambiguity.  In the presence of 
the six-body coupling of Eq.(\ref{sixbody}), three classes of loop 
diagrams can potentially contribute to the quartic self-coupling of 
the composite gauge boson in the leading 1/N order (Fig. 6). 
 
\begin{figure}[ht]
\epsfig{file=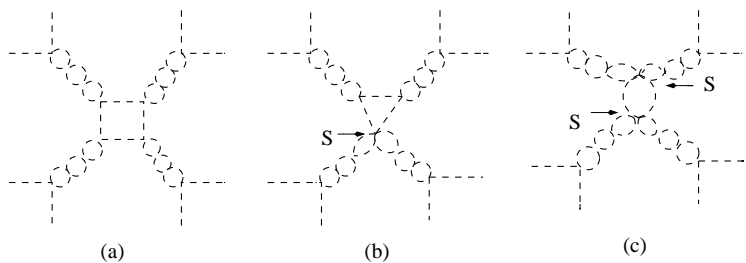,width=0.60\textwidth}
\caption{Three classes of diagrams can contribute to the quartic self-coupling 
of composite $A_{\mu}$. The letter $S$ for the six-body $\phi^*/\phi$ 
interaction point in the loop at center denotes that the $\phi^*\phi$ pair 
is in the scalar state. (a) No six-body coupling, (b) one six-body coupling,
and (c) two six-body couplings. \label{fig:6}}
\end{figure}

The square box diagrams (6a) alone do not cancel among themselves. When 
we add all three classes of the diagrams, however, they sum up to zero 
at the zero external momentum limit where the on-shell quartic coupling 
constant is defined.  Up to an overall constant, the cancellation occurs 
among the three types of diagrams in Fig. 6 as 
\begin{equation}
   \propto \Big(\frac{1}{4} - \frac{1}{2} +\frac{1}{4}\Bigr)
        \frac{1}{\Gamma(2-D/2)},
\end{equation}
where the first, second and third terms in the bracket are from the 
three types of diagrams, Figs. 6a, 6b and 6c, respectively.  Of course, 
this cancellation is not an accident.  Its origin is traced back to
the U(1) gauge invariance incorporated in the Lagrangian.\footnote{We 
freely switch between $\phi^*\phi$ and $:\!\phi^*\phi\!:$ in this  
calculation since our computation of the couplings 
involves only those diagrams in which a $\phi/\phi^*$ particle emitted from 
one $L_{int}$ annihilates at a \underline{another} $L_{int}$
in the center of diagram. See Figs. 6b and 6c. The normal ordering 
makes no difference in Figs. 6b nor 6c for this reason.} 

Our fundamental Lagrangian $L_{tot}$ is invariant under
the gauge transformation $\phi(x)\rightarrow e^{i\alpha(x)}\phi(x)$ and the 
conjugate. Once a massless vector bound-state emerges with the effective 
coupling $ie(\phi^*\stackrel{\leftrightarrow}{\partial}_{\mu}\phi) A^{\mu}$, 
the only way for it to be compatible with the gauge invariance is that the 
additional interaction $e^2\phi^*\phi A_{\mu}A^{\mu}$ exists for this effective 
$A_{\mu}$ field and that $A_{\mu}$ transforms as $eA_{\mu}\rightarrow eA_{\mu}+
i\partial_{\mu}\alpha$ under $\phi(x)\rightarrow e^{i\alpha(x)}\phi(x)$. 
As far as the interactions of dimension four are concerned, there is no 
other way known to us that satisfies the U(1) gauge invariance incorporated 
in $L_{tot}$.  As for the self-couplings of $A_{\mu}$, we would have to 
satisfy U(1) gauge invariance with the $A_{\mu}$ fields alone without
derivatives.  That is, there is no room to accommodate nonderivative 
self-interaction of $A_{\mu}$ in dimension four. When we argue in this way, 
gauge invariance of the composite $A_{\mu}$ coupling is an inevitable and 
trivial consequence of the gauge symmetry of $L_{tot}$, once a massless 
spin-one bound-state emerges with the coupling 
$ie(\phi^*\stackrel{\leftrightarrow}{\partial}_{\mu}\phi)A^{\mu}$. When 
we take this viewpoint, the crucial step is whether or not a massless 
bound state of spin-one is indeed formed out of the interactions among 
the matter fields themselves.  The rest may be interpreted as logical 
inevitability.

   Before closing this subsection, we comment on the interactions of 
dimension higher than four (in the world of space-time dimension four or
3+1).  The interaction $(\phi^*\phi)^2A_{\mu}A^{\mu}$ has dimension six.  
It can arise from the third term ($n=2$) of the expansion of the 
denominator $1/(\phi^*\phi)$ in Eq.(\ref{Lbint}), that is, 
\begin{equation}
   L_{int} = \frac{1}{4(I_0^b)^3}(:\phi^*\phi:)^2
      (\phi^*\stackrel{\leftrightarrow}{\partial}_{\mu}\phi)(
      \phi^*\stackrel{\leftrightarrow}{\partial}^{\mu}\phi).  \label{dim63}
\end{equation}
By attaching the chains of the $\phi$ bubbles to 
$(\phi^*\stackrel{\leftrightarrow}{\partial}_{\mu}\phi)$ and
$(\phi^*\stackrel{\leftrightarrow}{\partial}^{\mu}\phi)$, then 
going to the gauge-boson mass shells on the chains, we can extract 
the effective interaction of dimension six for the composite gauge boson,
\begin{equation}
   L_{int} = \frac{e^2}{I^b_0}(\phi^*\phi)^2A_{\mu}A^{\mu},    \label{dim61}
\end{equation} 
where the coupling $e^2$ is given by Eq.(\ref{U1bcoupling}). 
This coupling is not gauge invariant by itself. However, there is another
effective coupling of dimension six, which contains only a single $A_{\mu}$.
We can compute it with the interaction of Eq.(\ref{dim63}) and put it in the 
form of effective interaction,
\begin{equation}
   L_{int} = \frac{ie}{I_0^b}(\phi^*\phi)(
  \phi^*\stackrel{\leftrightarrow}{\partial}_{\mu}\phi) A^{\mu}. \label{dim62}
\end{equation}
When the two interactions Eqs.(\ref{dim61}) and (\ref{dim62}) of dimension six
are combined and added to the first term of the expansion of $1/(\phi^*\phi)$, 
\begin{equation}
       \frac{1}{4I_0^b}(\phi^*\stackrel{\leftrightarrow}{\partial}_{\mu}\phi)
                       (\phi^*\stackrel{\leftrightarrow}{\partial}^{\mu}\phi),
                                                        \label{dim60}
\end{equation}
the sum total is gauge invariant.  That is, when all the couplings of 
$O(1/I_0^b)$, Eqs.(\ref{dim61}), (\ref{dim62}) and (\ref{dim60}) are combined, 
the interaction of dimension six for the effective field $A_{\mu}$ is 
gauge invariant.  The combined effective interaction can be cast into the form 
\begin{equation}
  L_{int}^{eff} = \frac{1}{4I_0^b}(\phi^*\stackrel{\leftrightarrow}{D}^{\mu}\phi)
            (\phi^*\stackrel{\leftrightarrow}{D}_{\mu}\phi), \label{eff}
\end{equation}
where $D_{\mu}=\partial_{\mu}+ieA_{\mu}$ and 
$(\phi^*\stackrel{\leftrightarrow}{D}_{\mu}\phi) \equiv 
\phi^*D^{\mu}\phi - (D_{\mu}\phi)^*\phi$.  
The interaction of Eq.(\ref{eff}) illustrates what happens for the effective 
interactions of higher dimension in general.  It is obvious from the dimensional 
reason that $L_{int}^{eff}$ must be inversely proportional to powers of $I_0^b$.   
Although $I_0^b$ is formally proportional to $m^2$ in the dimensional
regularization, it is quadratically divergent in the cutoff 
($\sim N\overline{\Lambda}^2$) in the world of $D=4$.  If we give a physical 
meaning to the cutoff, therefore, the interactions of dimension six are 
suppressed by $O(p^2/N\overline{\Lambda}^2)$ in the region of energy scale 
$O(p^2)$ relative to those of dimension four.  Meanwhile the divergences of 
$O(N\ln\overline{\Lambda}^2)$ are absorbed into the gauge coupling $e^2$ 
as we have seen in Eq.(\ref{U1bcoupling}).   
Therefore, if our model should turn out to be phenomenologically relevant 
in one way or another, its cutoff $\overline{\Lambda}$ would place these 
higher-dimensional interaction under control. Whether these interactions 
can generate anything phenomenologically interesting or not is a separate 
question.

We can cast the amplitudes of higher-dimension processes in the standard
U(1)-gauge theory with the elementary gauge boson into the form of effective 
interactions. However, such
effective interactions are generally not identical with the higher 
dimensional interactions that have been obtained above from our Lagrangian 
Eq.(\ref{Lb}).  The loop-diagram amplitudes produced by the standard U(1) 
gauge theory do exist equally in our model since the gauge boson exists 
as a composite.  Our model contains the additional terms that are generated 
by matter fields and suppressed by the large cutoff scale of $I_b$. Physics 
is generally different in these orders from the standard gauge theory of 
elementary gauge boson. If our model were identical with the standard U(1) 
gauge theory, it would be perfectly renormalizable in our world of dimension 
four.  But that is not the case: Our model contains the higher-dimensional 
local interactions that are additional to the standard gauge theory and 
suppressed by powers of $1/I_b = O(\overline{\Lambda}^2)$.
  
\subsection{Gauge boson in fermionic model}

Computation of the massless bound state is technically a little complex 
in the fermionic model since there exist two channels of $J^{PC}=1^{--}$.
We compute the elastic scattering of fermion-antifermion,
\begin{equation}
    f^+(p_1,s_1)+ f^-(p_2,s_2) \rightarrow f^+(p_3,s_3) + f^-(p_4,s_4)
                                   \label{Fscat}
\end{equation}
in the leading 1/N order with the Lagrangian Eq.(\ref{Lf}).  The copy indices 
are chosen to be the same for the initial $f^+f^-$ and for the final $f^+f^-$.  
We shall suppress spin indices $s_i (i=1,\cdots 4)$ in the following since they 
are obvious in most places.  We leave out the {\em self-gauge invariant} 
interactions such as Eq.(\ref{Lfprime}).  Although those interactions certainly 
contribute to the fermion-fermion scattering in general, we show later that 
omission of such interactions does not affect the properties of the massless 
bound state.

We follow our path taken for the bosonic model: We separate $\overline{\psi}\psi$ 
in the denominator of $L_{int}$ into sum of the vacuum expectation values and 
the normal-ordered products $:\overline{\psi}\psi:$ and then expand it in the
power series of $\sum :\overline{\psi}\psi:/\sum 
\langle 0|\overline{\psi}\psi|0\rangle$. The vacuum expectation value  
$\langle 0|\overline{\psi}\psi|0\rangle$ is divergent and 
dimensionally regularized as
\begin{eqnarray}
 \sum\langle 0|\overline{\psi}\psi|0\rangle &=& -\lim_{x\rightarrow 0}
              {\rm tr}\langle 0|T(\psi(x)\overline{\psi}(0)|0\rangle , 
                                       \nonumber \\
      &=& -\frac{4Nm\Gamma(1-D/2)}{(4\pi)^{D/2}(m^2)^{1-D/2}}, \label{vac1f}
\end{eqnarray}
where the trace (tr) in the first line of the right-hand side refers to the
spinor indices of $\psi$ and $\overline{\psi}$.
We shall denote the right-hand side of Eq.(\ref{vac1f}) by $I^f_0$ hereafter as
\begin{equation}
   I^f_0 \equiv \langle 0|\overline{\psi}\psi|0\rangle  = -4mI^b_0.    \label{vac2f}
\end{equation}
$I_0^f$ is opposite in sign to $I^b_0$ of the boson Eq.(\ref{vac1}) and
its dimension is three instead of two.

Now we proceed to compute the two-body scattering amplitude of $J^{PC}=1^{--}$.
There exist two eigenchannels in the fermion scattering. The fermion-antifermion 
pair is in the configuration of $\overline{v}_{-\bf p}\mbox{\boldmath$\gamma$}u_{\bf p}$ 
in one channel and in $2{\bf p}\overline{v}_{-\bf p}u_{\bf p}$ in the other in 
the center-of-momentum frame. The spins of $\overline{v}_{-\bf p}$ and $u_{\bf p}$ 
are combined into a triplet in both cases so that they make the $^3S_1$ 
and $^3D_1$ states of $f^+f^-$, respectively.  With our choice of $L_{int}$ 
in Eq.(\ref{Lf}), the fermion-antifermion pair turns from 
$\overline{\psi}\gamma_{\mu}\psi$ on one side to 
$(\overline{\psi}\stackrel{\leftrightarrow}{\partial}^{\mu}\psi)$ 
on the other, or conversely from 
$(\overline{\psi}\stackrel{\leftrightarrow}{\partial}_{\mu}\psi)$ to 
$\overline{\psi}\gamma^{\mu}\psi$ at every interaction point in the chain 
of bubbles. 

Let us define the Lorentz scalar amplitude $T(p_1,p_2;p_3,p_4)$ 
with the S-matrix as
\begin{equation}
     <p_3,p_4|S-1|p_1,p_2> =i(2\pi)^4 \delta^4(p_3+p_4-p_1-p_2)
                     T(p_3,p_4;p_1,p_2),  \label{Tamp}
\end{equation}
where the one-fermion states are so normalized that the amplitude 
$T(p_3,p_4;p_1,p_2)$ is a Lorentz scalar and its Lorentz structure 
is given in the ($2\times 2$) matrix form by
\begin{equation}
 T = \left(\begin{array}{cc} \overline{u}_{p_3}\gamma_{\mu}v_{p_4},
                             \overline{u}_{p_3}(p_3-p_4)_{\mu}v_{p_4}/m 
                                     \end{array} \right)
     \left(\begin{array}{cc}   T_{11}^{\mu\nu}(q) & T_{12}^{\mu\nu}(q)\\
                               T_{21}^{\mu\nu}(q) & T_{22}^{\mu\nu}(q)
                                     \end{array} \right)
     \left(\begin{array}{c}
             \overline{v}_{p_2}\gamma_{\nu}u_{p_1} \\
             \overline{v}_{p_2}(p_1-p_2)_{\nu}u_{p_1}/m
                                     \end{array} \right),
                                            \label{amplitudef}
\end{equation}
where $q=(p_1+p_2)=(p_3+q_4)$.  The perturbation series for $T(q)^{\mu\nu}$ 
starts with the tree diagram, which gives $-(\lambda/2I_0^f)g_{\mu\nu}$ to 
the off-diagonal elements of $T^0_{\mu\nu}$:
\begin{equation}
     T^0_{\mu\nu}  = -\frac{1}{2I^f_0}
                        \left( \begin{array}{ll}
                               0  &  \lambda          \\
                               \lambda  &  0
                        \end{array}  \right)
                              g_{\mu\nu}  .
                             \label{lowest}
\end{equation} 
Summation of the bubble chains can be carried out by solving the
matrix equation, 
\begin{equation}
 T(q)_{\mu\nu} =T^0_{\mu\nu} + K(q)_{\mu\kappa}T^{\kappa}_{\nu}(q), 
                                        \label{AlEq}
\end{equation}
where the kernel $K(q)_{\mu\kappa}$ is the $2\times 2$ matrix of the 
four single-bubble diagrams that connect between $\gamma_{\mu}$-type vertex 
($^3S_1$) and the $\stackrel{\leftrightarrow}{\partial}_{\mu}$-type 
vertex ($^3D_1$). (See Fig. 7.)

\begin{equation}
 K^{\mu\kappa}(q) =  \left( \begin{array}{cc} 
                     K(q)_{11} & K(q)_{12} \\
                     K(q)_{21} & K(q)_{22}  
                    \end{array}  \right)^{\mu\kappa}.       \label{Kmatrix}
\end{equation}

\noindent
\begin{figure}[ht]
\epsfig{file=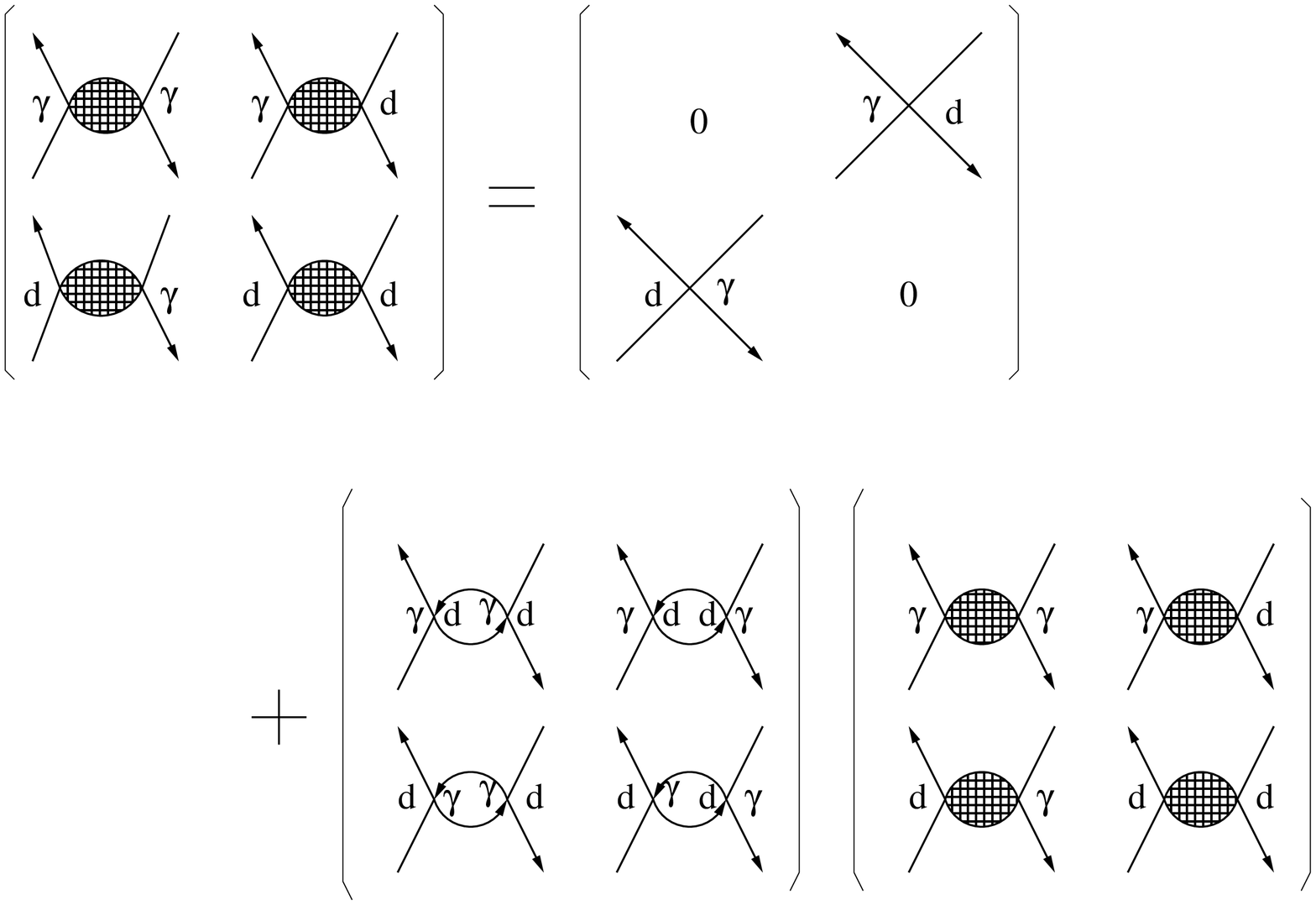,width=0.47\textwidth}
\caption{Iteration of bubble diagrams for fermion scattering. The letters 
$\gamma$ and $d$ denote that the fermion pair at the interaction point is 
$\overline{\psi}\gamma_{\mu}\psi$ and $\overline{\psi}\stackrel{
\leftrightarrow}{\partial}_{\mu}\psi$, respectively.
\label{fig:7}}
\end{figure}
  
In order to extract the mass and coupling of the composite boson from 
$T(q)_{\mu\nu}$, we need $(I-K(q))_{\mu\kappa}$ near $q^2=0$ in 
Eq.(\ref{AlEq}). To be more specific, the terms of $g_{\mu\kappa}$ and 
$(q^2 g_{\mu\kappa}-q_{\mu}q_{\kappa})$ for $K_{ij}$. In fact, for the 
off-diagonal elements $K_{12}$ and $K_{21}$, all we need is the leading 
terms that give $K_{12}K_{21}=O(q^2)$. By straightforward diagram 
computation, we find the relevant terms of $K^{\mu\kappa}(q)$ near 
$q^2 = 0$ as
\begin{eqnarray}
   K^{\mu\kappa}(q)_{11} &=& \lambda\Bigl(g^{\mu\kappa}
            +\frac{\Gamma(2-D/2)}{6m^2\Gamma(1-D/2)} 
                      (g^{\mu\kappa}q^2-q^{\mu}q^{\kappa})\Bigr),
                                        \nonumber \\
                      &=&  K^{\mu\kappa}(q)_{22}  \nonumber \\    
   K^{\mu\kappa}(q)_{12} &=& -\lambda \Bigl(\frac{\Gamma(-D/2)}{
        \Gamma(1-D/2)}-2\Bigr)g^{\mu\kappa}, \nonumber \\
   K^{\mu\kappa}(q)_{21} 
         &=& -\lambda\Bigl(\frac{\Gamma(2-D/2)}{6m^2\Gamma(1-D/2)}\Bigr)
           (g^{\mu\kappa}q^2-q^{\mu}q^{\kappa}).  \label{Kelement}
\end{eqnarray}
We have kept $\Gamma$-functions above since they are partially canceled 
with $\Gamma(1-D/2)$ coming from $1/I_0^f$ of $T^0$ when $(I-K)^{-1}$ 
is operated on $T^0$ later.  The terms in Eq.(\ref{Kelement}) that 
turn out to determine the pole and residue of the massless bound state 
are the first term $\lambda g^{\mu\kappa}$ of the diagonal element 
$K(q)^{\mu\kappa}_{11} (= K(q)^{\mu\kappa}_{22})$  and the
off-diagonal element $K(q)^{\mu\kappa}_{12}\neq 0$ at $q^2=0$.  

Let us examine the pole and residue of the matrix amplitude $T_{\mu\nu}$ 
at $q^2=0$ by solving Eq.(\ref{AlEq}) as
\begin{equation}
   T_{\mu\nu}  = \Bigl(\frac{1}{I-K}\Bigr)_{\mu}^{\kappa}T^0_{\kappa\nu}.      
                                           \label{solution}
\end{equation}
Since the external fermion lines are on mass shell, the terms proportional 
to $q_{\mu}q_{\kappa}$ in $K_{\mu\kappa}$ has been removed by use of the Dirac 
equation and the mass shell condition on the external lines. 
We then approach the gauge symmetry limit $\lambda = 1$ of 
$T= (I-K)^{-1}T^0$.  The result is
\begin{equation}
   T(q)_{\mu\nu} = \frac{(4\pi)^{D/2}(m^2)^{2-D/2}}{\Gamma(2-D/2)}
                   \left(  \begin{array}{cc}
                  \frac{3}{4q^2}   & \frac{C}{m^2} \\
                  \frac{C}{m^2} & 
                 \frac{C}{m^2} 
                   \end{array} \right)g_{\mu\nu},  \label{sol}
\end{equation}
where 
\begin{equation}
      C = \frac{D(D-2)}{32(D+1)}. \label{constant}
\end{equation}
A pole appears only in the $(11)$-matrix element at the upper left corner 
in Eq.(\ref{sol}) and the other entries are regular at $q^2=0$.  This is 
depicted in Fig. 8.

\noindent
\begin{figure}[ht]
\epsfig{file=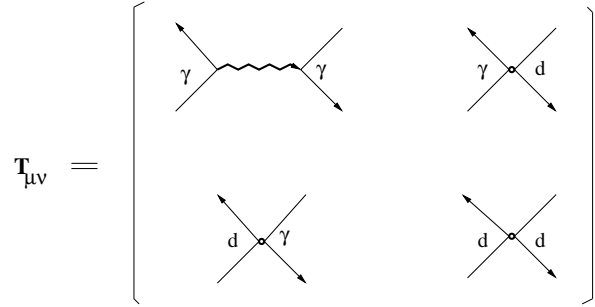,width=0.47\textwidth}
\caption{The massless bound state appears only in the upper left corner,
which is the $^3S_1$ channel.
\label{fig:8}}
\end{figure}

It means that bound state appears in the channel of 
$\overline{\psi}\gamma_{\mu}\psi\rightarrow \overline{\psi}\gamma_{\mu}\psi$, 
that is, in the $^3S_1$ channel, not in the $^3D_1$ channel.\footnote{This 
has nothing to do with the $d$-wave threshold behavior $\sim |{\bf p}|^l (l=2)$. 
The threshold behaviors reside in the spinorial factors in Eq.(\ref{amplitudef}) 
and have been separated out in defining $T(q)_{ij}^{\mu\nu}$.}  If either end of
the chain is $\overline{\psi}\stackrel{\leftrightarrow}{\partial}_{\mu}\psi$, 
no massless pole appears in such a chain.

By comparing the matrix element $T_{11}^{\mu\nu}$ with the one-photon 
pole diagram of the standard U(1) gauge interaction 
$-e\overline{\psi}\gamma_{\mu}\psi A^{\mu}$, we can identify the gauge 
coupling $e^2$ with the residue at the pole to obtain
\begin{equation}
  e^2 = \frac{3(4\pi)^{D/2}(m^2)^{(2-D/2)}}{4N\Gamma(2-D/2)}. \label{U1fcoupling0}
\end{equation}
or in terms of the covariant ultraviolet cutoff in the space-time of $D=4$,
\begin{equation}
      e^2 = \frac{12\pi^2}{N\ln(\overline{\Lambda}^2/m^2)}. 
                                        \label{U1fcoupling}
\end{equation}
This is the parallel of Eq.(\ref{U1bcoupling}) in the bosonic model. 
While the quartic divergence ($\propto \Gamma(-D/2) \sim \Lambda^4$) 
and quadratic divergence ($\sim \Lambda^2$) are present in $T(q)_{\mu\nu}$, 
they do not enter the residue of the pole at $q^2=0$.  Therefore, 
the coupling $e^2$ involves only the logarithmic divergence 
($\sim 1/N\ln\Lambda^2$) as it does for the bosonic model. 

  As we have pointed out, we may add to our fermionic model the interaction 
$L'_{int}$ of Eq.(\ref{Lfprime}) which is gauge invariant by itself.  Let 
us denote the shifts of the matrices $K(q)$ and $T^0$ due to $L'_{int}$ as 
$K(q)\rightarrow K(q)+\Delta K(q)$ and $T^0\rightarrow T^0+\Delta T^0$.  
Near $q^2=0$, these shifts are given by 
\begin{equation}
           \Delta T^0_{\mu\nu} =  \frac{1}{2I_0^f}g_{\mu\nu}      
           \left( \begin{array}{cc}
                   fm &   0 \\   
                   0  &     0   
          \end{array} \right). \label{mod1}
\end{equation}
and
\begin{equation}
   \Delta K^{\mu\kappa} = \frac{1-D/2}{6m^2}        
           \left( \begin{array}{cc}
                   f  &  0 \\   
                   0  &     0   
          \end{array} \right)(g^{\mu\kappa}q^2-q^{\mu}q^{\kappa}).  \label{mod2}
\end{equation}
It is not difficult to see that these modifications, Eqs.(\ref{mod1}) and 
(\ref{mod2}), do not alter either the location of the pole at $q^2=0$ nor 
its residue.   In terms of diagrams, 
we can visualize the effect of Eqs.(\ref{mod1}) and (\ref{mod2}) as follows: 
We should first notice the fact that the newly added bubble consisting of 
$\gamma_{\mu}$ on one end and $\gamma_{\kappa}$ on the other end vanishes 
like $(g_{\mu\kappa}q^2-q_{\mu}q_{\kappa})$ at $q = 0$.  Let us call 
this bubble as that of the type $\gamma_{\mu}\otimes\gamma_{\kappa}$.  
When the $\gamma_{\mu}\otimes\gamma_{\kappa}$ bubble enters the middle of 
the eigenchannel that produces the bound state, the chain would thus acquire 
a factor of $O(q^2)$ from this bubble.  Therefore it cancels the pole and 
becomes irrelevant to formation of the massless bound state. The pole at 
$q^2=0$ is produced only by the $g_{\mu\kappa}$ term of $K(q)_{\mu\kappa}$ 
in the chain of bubbles of the type 
$\gamma_{\mu}\otimes\stackrel{\leftrightarrow}{\partial}_{\nu}$ and
$\stackrel{\leftrightarrow}{\partial}_{\mu}\otimes\gamma_{\nu}$ alone.
With the addition of $L'_{int}$, therefore, the massless pole is
undisturbed and its residue is unaffected.     

Let us move on to the self-coupling of the gauge field.  Charge conjugation 
invariance forbids the triple self-coupling, but the quartic self-coupling is 
not forbidden by any discrete symmetry. Since the massless bound state couples 
only to the $^3S_1$ vertex, namely, to $\overline{\psi}\gamma_{\mu}\psi$, 
the relevant diagrams have a square box at the center
with six permutations of the four $\gamma$-vertices, that is, the diagram of
Fig 6a in which the boson lines are replaced by the fermion lines and the
$\gamma$-matrices sit at the four corners of the box. However, the sum of these 
box diagrams vanishes in the zero-energy-momentum limit of the bound-state  
bosons, not just the leading divergent term ($\sim \ln\overline{\Lambda}^2$)
but all finite terms as well in this limit. This fact is well-known 
as the gauge-invariance requirement $\sim e^4 F_{\mu}^{\nu}F_{\nu}^{\kappa}
F_{\kappa}^{\lambda}F_{\lambda}^{\mu}$ on the photon-photon scattering
amplitude in quantum electrodynamics.  

For the diagrams corresponding to Fig 6b and 6c with the boson lines replaced 
by fermions, the two chains of bubbles are attached to the six-body fermion 
interaction. However, since the six-body fermion interaction is of the form 
$(\overline{\psi}\gamma_{\mu}\psi)(\overline{\psi}\stackrel{\leftrightarrow}{
\partial}^{\mu}\psi)(\overline{\psi}{\psi})$, one of the vector vertices 
starts with the $\gamma$-vertex but the other starts with the 
$\stackrel{\leftrightarrow}{\partial}$-vertex. As we have already observed,
the massless bound-state pole cannot appear in the latter chain. Therefore, 
the massless vector bound state can be formed only in one of the two chains 
attached to the six-body interaction point, not in both. That is, only three 
massless bound states can be formed in Fig. 6b and two in Fig. 6c.  
Combining this observation with that for Fig. 6a above, we conclude that 
there exists no nonderivative quartic self-coupling of the massless U(1) 
bound-state in the fermion model either, just as gauge invariance requires.

The lowest possible couplings of higher dimension with fermion fields is 
the Pauli term $i\overline{\psi}\sigma_{\mu\nu}\psi F^{\mu\nu}$. This coupling 
is gauge invariant by itself. With our interaction $L_{int}$, however, 
our composite boson does not have this coupling. To see this, recall 
the decomposition of the photon-fermion vertex for the fermion on mass shell, 
$i\overline{u}\sigma_{\mu\nu}q^{\nu}v' = \overline{u}(p+p')_{\mu}v'- 2m
\overline{u}\gamma_{\mu}v'$.  This relation tells that if the massless bound 
state had the Pauli-term interaction, we would have its pole in the channels 
of both $\overline{\psi}\stackrel{\leftrightarrow}{\partial}_{\mu}\psi$ and
$\overline{\psi}\gamma_{\mu}\psi$. In our preceding study, however, we have 
found a massless pole only in $\overline{\psi}\gamma_{\mu}\psi$. That means 
no Pauli term.

The effective interaction $\overline{\psi}\psi A_{\mu}A^{\mu}$ is also of 
dimension five and not gauge invariant by itself.  As in the bosonic model, 
If an interactions of $A_{\mu}$ appears with a dimension higher than four, 
it ought to appear in a gauge invariant combination since the underlying 
Lagrangian is gauge invariant. As for this specific interaction, the 
accompanying gauge-covariant partners are 
$\partial^{\mu}\overline{\psi}\partial_{\mu}\psi$ and 
$ie(\overline{\psi}\stackrel{\leftrightarrow}{\partial}_{\mu}\psi)A^{\mu}$.
But we have already found that the coupling $(\overline{\psi}
\stackrel{\leftrightarrow}{\partial}_{\mu}\psi)A^{\mu}$ does not exist in 
our model. Neither $\partial_{\mu}\psi\partial^{\mu}\psi$ in $L_{tot}$. 
Therefore the coupling $(\overline{\psi}\psi)A_{\mu}A^{\mu}$ can be 
generated as an effective interaction in our model.

One of the merits of our fermionic model is to reveal the dynamical details
explicitly in regard to how the self-interaction of the constituent fermions 
conspires to generate the composite gauge boson. Specifically, the composite
gauge boson is formed with fermions in the presence of the process of 
the transition between the $^3S_1$ and the $^3D_1$ channel. No massless 
bound state can be formed with the $^3S_1$  channel alone.  There is no 
place to see this dynamics in the auxiliary field trick on fermions
in which the auxiliary vector field has only the $^3S_1$ interaction.

\section{Non-Abelian extensions}
  It is possible to extend our U(1) models to non-Abelian models.  
The non-Abelian extension turns out to be quite easy if we choose 
matter fields in the SU(2) doublet. In this section we present the 
SU(2)-doublet model for both bosons and for fermions and compute for 
the composite gauge bosons again in the leading 1/N order. Extension of 
our U(1) models to a general Lie group or even to an SU(2) representation  
other than the doublet encounters difficulty. This is not a simple 
technical difficulty, but it involves some problem at a fundamental 
level in our class of models. We explain this difficulty in the text, 
then go a little further with few examples of the bosonic models 
in Appendix B.

Those who approach the problem with the auxiliary field trick would 
trivially extend the U(1) model to general groups and representations
by simply replacing the two-by-two matrices $\frac{1}{2}\tau_a$ of SU(2) 
with the $n\times n$ generator matrices $T_a$ of a general Lie group. 
In our case, however, such simple substitution does not extend our models 
to those of general groups or representations.\footnote{One well-known 
example of the special role of the SU(2) may come to mind, i.e., 
the instanton. The instanton is special to SU(2), not extendable 
to SU(N) $(N \geq 3)$ or other general groups because of its topological 
property.  In our case, however, topology is not an issue.  Important is the
{\em self-duality} of the group and the representation.} This is another 
indication of the fact that our models are physically different at some
fundamental level from what the auxiliary field trick gives.

\subsection{Non-Abelian bosonic model}

Let us introduce $N$ families of scalar boson fields in SU(2) doublet,
\begin{equation}
   \Phi^{i} = \left(  \begin{array}{c}
                 \phi_1^i \\
                 \phi_2^i   \end{array} \right),
                 \;\;\;(i= 1, \cdots N),        \\
                   \label{bdoublet}
\end{equation} 
and their conjugates $\Phi^{i\dagger}$, which we write in a row.   The
subscripts $(1,2)$ are those of SU(2). We shall suppress the copy index and/or 
the SU(2) index wherever there is no confusion.  Our bosonic Lagrangian 
is given simply by
\begin{eqnarray}
 L_0 &=&\sum_i \partial^{\mu}\Phi^{i\dagger}\partial_{\mu}\Phi^i
                             - \sum_i m^2\Phi^{i\dagger}\Phi^i \nonumber \\
 L_{int}&=& \lambda \frac{\sum_i
   (\Phi^{i\dagger} \mbox{\boldmath $\tau$}\stackrel{\leftrightarrow}{\partial}_{\mu}\Phi^i)\cdot
    \sum_j   
   (\Phi^{j\dagger} \mbox{\boldmath $\tau$}\stackrel{\leftrightarrow}{\partial}^{\mu}\Phi^j)}{
    4\sum_k (\Phi^{k\dagger}\Phi^k)},  \;\;(\lambda\rightarrow 1)        \label{NAb}
\end{eqnarray}
where $i$, $j$ and $k$ are copy indices and $\mbox{\boldmath $\tau$}$ denotes 
the Pauli matrices $\tau_a (a=1,2,3)$.\footnote{This bosonic Lagrangian as well as 
its Abelian version appears in the earlier paper{\cite{Akh}}.}  For the SU(2) gauge 
invariance of $L_0+L_{int}$, we give the proof here for the infinitesimal rotation,
\begin{eqnarray}
     \Phi &\rightarrow& (1+\frac{i}{2}\mbox{\boldmath$\tau\cdot\alpha$})\Phi, \nonumber \\
     \Phi^{\dagger} &\rightarrow& \Phi^{\dagger} (1-\frac{i}{2}\mbox{\boldmath$\tau\cdot\alpha$}), 
                          \label{brot}
\end{eqnarray}
where $\mbox{\boldmath$\alpha$}$ is a space-time dependent vector function.
Let us compute the variations $L_0\rightarrow L_0+\delta L_0$ and 
$L_{int}\rightarrow L_{int}+ \delta L_{int}$ separately and confirm cancellation 
to $O(\alpha)$ between the two variations.  For $L_0$, it is easy to obtain 
\begin{equation}
    \delta L_0 = 
 -\frac{i}{2}(\Phi^{\dagger}\mbox{\boldmath$\tau$}\stackrel{\leftrightarrow}{\partial}_{\mu}
     \Phi)\cdot\partial^{\mu}\mbox{\boldmath$\alpha$} +O(\alpha^2). 
                              \label{varL0}
\end{equation}
We need a little care in computation of $\delta L_{int}$.  To the order $O(\alpha)$, 
it is not difficult to obtain the transformation,
\begin{equation}
 (\Phi^{\dagger}\mbox{\boldmath $\tau$}\stackrel{\leftrightarrow}{\partial}^{\mu}\Phi) 
  \rightarrow\Phi^{\dagger}U^{\dagger}\mbox{\boldmath $\tau$} U\partial^{\mu}\Phi 
       -(\partial^{\mu}\Phi^{\dagger})U^{\dagger}\mbox{\boldmath $\tau$} U\Phi 
       +2i(\Phi^{\dagger}\Phi)
                   \partial^{\mu}\mbox{\boldmath $\alpha$} + O(\alpha^2), \label{calcu}
\end{equation}
where $U=1+i\mbox{\boldmath $\tau$}\cdot\mbox{\boldmath $\alpha$}/2$. 
The third term proportional to $\partial^{\mu}\mbox{\boldmath $\alpha$}$ 
in the right-hand side has been obtained by use of the relation, 
\begin{equation}
        \mbox{\boldmath$\tau$}(\mbox{\boldmath$\tau$}\cdot
        \partial^{\mu}\mbox{\boldmath$\alpha$}) +
        (\mbox{\boldmath$\tau$}\cdot
        \partial^{\mu}\mbox{\boldmath$\alpha$})\mbox{\boldmath$\tau$}
         = 2\partial^{\mu}\mbox{\boldmath$\alpha$}.   \label{tau}
\end{equation}
Since an isoscalar product remains unchanged under global SU(2) rotations, 
it holds for arbitrary SU(2)-doublet functions, $A, B, C$ and $D$, that 
\begin{equation} 
\Big((UA)^{\dagger}\mbox{\boldmath $\tau$}UB\Big)\cdot
    \Big((UC)^{\dagger}\mbox{\boldmath $\tau$}UD\Big)
  =(A^{\dagger}\mbox{\boldmath $\tau$}B)\cdot(C^{\dagger}\mbox{\boldmath $\tau$}D).
\end{equation}
Thanks to this relation, when we take product of Eq.(\ref{calcu}) with 
itself in $L_{int}$, four products made of the first two terms are invariant 
by themselves as 
\begin{equation}
     (\Phi^{\dagger}U^{\dagger}\mbox{\boldmath$\tau$}U\partial_{\mu}\Phi)\cdot
   (\Phi^{\dagger}U^{\dagger}\mbox{\boldmath$\tau$}U\partial^{\mu}\Phi) =
              (\Phi^{\dagger}\mbox{\boldmath$\tau$}\partial_{\mu}\Phi)\cdot
              (\Phi^{\dagger}\mbox{\boldmath$\tau$}\partial^{\mu}\Phi),    
\end{equation}
and so forth. The product of the third term with itself is $O(\alpha^2)$. 
In the cross products of the first two terms with the third term 
$2i(\Phi^{\dagger}\Phi)\partial^{\mu}\mbox{\boldmath $\alpha$}$, we may set 
$U=1$ since we are computing to $O(\alpha)$. Dividing these terms of $O(\alpha)$
in the numerator of $\delta L_{int}$ by $4(\Phi^{\dagger}\Phi)$, 
we obtain that the variation of $L_{int}$ is equal to 
\begin{equation} 
 +\frac{i}{2}\lambda(\Phi^{\dagger}\mbox{\boldmath$\tau$}\stackrel{\leftrightarrow}{
    \partial}_{\mu}\Phi)\cdot\partial^{\mu}\mbox{\boldmath$\alpha$}+O(\alpha^2),
\end{equation}
which cancels $\delta L_0$ for $\lambda = 1$. 

The proof to all orders of $\mbox{\boldmath $\alpha$}$ is not difficult 
though a bit tedious.  We can carry it out with brute force using the local 
rotation matrix $U$ for the SU(2) doublet matter fields,   
\begin{equation}
    U = \cos\alpha 
    + i(\mbox{\boldmath$\hat{\alpha}$}\cdot\mbox{\boldmath$\tau$})\sin\alpha,
                            \label{finite}
\end{equation}
where $\mbox{\boldmath${\hat\alpha}$}=\mbox{\boldmath$\alpha$}/\alpha$.
Alternatively, in the case of bosons, we could introduce the auxiliary fields 
and integrate over them to reach the Lagrangian Eq.(\ref{NAb}).  Operationally, 
this turns out to be a much simpler avenue.  While its physical meaning is 
subject to debate or some people feel it questionable, we can use the auxiliary 
field method as a mathematical tool of manipulation without a problem. If one 
wants to proceed along that line, one starts with 
\begin{equation}
    L_{tot} = \bigl(\partial^{\mu}+i\mbox{\boldmath $A$}^{\mu}\Phi\bigr)^{\dagger}\cdot
 (\partial_{\mu}+i\mbox{\boldmath $A$}_{\mu})\Phi - m^2\Phi^{\dagger}\Phi 
  +\frac{1}{2}\mu^2\mbox{\boldmath $A$}^{\mu}\mbox{\boldmath $A$}_{\mu},
\end{equation}
where ${\bf A}^{\mu} = \frac{1}{2}\tau_aA_a^{\mu}$. Although we do not really 
need it here, we have added the mass term $\mu^2$ to ${\bf A}^{\mu}$ 
for gauge fixing, which is to be removed after functional interaction 
is completed.

 Having seen the Lagrangian of Eq.(\ref{NAb}), it is tempting to speculate 
that if the isospin $\frac{1}{2}\tau_a$ is replaced by the $n\times n$ 
matrices of the generator $T_a$ of some other group $G$, we could obtain 
the non-Abelian extension to the case where the matter fields form the 
$n$-dimensional multiplets of group $G$. Namely,
\begin{equation}
 L_{tot} =  \sum_i \partial^{\mu}\Phi^{i\dagger}\partial_{\mu}\Phi^i
                            - m^2\Phi^{i\dagger}\Phi^i +
    \lambda \frac{\sum_i
   (\Phi^{i\dagger} T_a\stackrel{\leftrightarrow}{\partial}_{\mu}\Phi^i)\cdot
    \sum_j   
   (\Phi^{j\dagger} T_a\stackrel{\leftrightarrow}{\partial}^{\mu}\Phi^j)}{
     \sum_k (\Phi^{k\dagger}\Phi^k)},  \;\;(\lambda\rightarrow 1),
     \label{NAbgeneral}
\end{equation}
where $T_a \neq \frac{1}{2}\tau_a$.  Unfortunately, this does not work.  
{\em The Lagrangian of Eq.(\ref{NAbgeneral}) is not gauge invariant}. 
We can pinpoint the step where the proof fails in this attempt: The relation 
of Eq.(\ref{tau}) is crucial in achieving non-Abelian gauge invariance in the  
Lagrangian Eq.(\ref{NAb}). This relation holds only for the SU(2) doublet. 

Some may yet wonder why one cannot resort to the auxiliary field trick 
starting with
\begin{equation}
    L_{tot} =  (\partial^{\mu}+i\mbox{\boldmath $A$}^{\mu}\Phi)^{\dagger}\cdot
 (\partial_{\mu}+i\mbox{\boldmath $A$}_{\mu})\Phi - m^2\Phi^{\dagger}\Phi,  
                                                       \label{NAbaux},                  
\end{equation}
where $\mbox{\boldmath A}_{\mu}=T_aA_{\mu}^a$.  The equation of motion for 
the auxiliary field $A_{\mu}^a$ is to be obtained by solving
\begin{equation}
  - i(\Phi^{\dagger}T_a\stackrel{\leftrightarrow}{\partial}_{\mu}\Phi)
   + \Phi^{\dagger}\{T_a,T_b\}\Phi A_{\mu}^b = 0.  \label{NAEofM}                       
\end{equation}
The $n\times n$ matrix $\{T_a,T_b\}$ is not proportional 
to a unit matrix except in the case of
$T_a = \frac{1}{2}\tau_a$.  In fact, its determinant is zero in most cases.
Consequently, the set of the algebraic equations Eq.(\ref{NAEofM}) is generally 
unsolvable. This same problem derails an attempt to integrate over the field 
$A_{\mu}^a$ to get an effective action in terms of $\Phi$ and $\Phi^{\dagger}$ 
alone.  We have illustrated this difficulty by two examples in Appendix B.

   When one attempts diagram calculation with the wrong Lagrangian of 
Eq.(\ref{NAbgeneral}), one could tune the location of a pole in the chain 
of the bubble diagrams to zero by setting $\lambda$ off unity.  
However, when one proceeds to calculate the coupling of 
$\Phi^{\dagger}\Phi A_{\mu}A^{\mu}$ (see Fig. 3), the Lagrangian of 
Eq.(\ref{NAbgeneral}) would generate the form
\begin{equation}
       \Phi^{\dagger}\Phi {\bf A}_{\mu}\cdot{\bf A}^{\mu},
\end{equation}
where the structure ${\bf A}_{\mu}\cdot{\bf A}^{\mu}$ arises from the 
denominator of $L_{int}$ and enters the center of the triangular loop 
in Fig. 3. However, the correct non-Abelian structure for these 
couplings ought to be 
\begin{equation}
       \Phi^{\dagger}\{T_a,T_b\}\Phi A_{\mu}^a A^{\mu b}.
\end{equation}
This conflict is another manifestation of the fact that the Lagrangian 
of Eq.(\ref{NAbgeneral}) is not gauge invariant.

These arguments are more than what we really need, but they hopefully 
clarify the special role of the SU(2) doublet matter fields when we 
attempt to write a {\em local} non-Abelian gauge invariant Lagrangian 
with matter fields alone. We have not succeeded in finding such 
a Lagrangian in a reasonably simple form except for the SU(2) doublet 
matters.    

\subsection{Non-Abelian fermionic model}

The non-Abelian extension is possible for the fermionic model if one 
follows the bosonic model given above.  For the SU(2) gauge group 
where the Dirac fields form SU(2) doublets with $N$ copies,
\begin{eqnarray}
        \Psi^i &=& \left( \begin{array}{c}   \psi^i_1  \\
                                           \psi^i_2   
                       \end{array} \right)         \nonumber \\
        \overline{\Psi}^i &=& (\overline{\psi}^i_1,\overline{\psi}^i_2), 
                    \;\; \mbox(i=1,2\cdots N),
                                              \label{fdoublet}
\end{eqnarray}
the gauge invariant Lagrangian is given by
\begin{eqnarray}
  L_0 &=&\sum_{i=1} \overline{\Psi}^i(i\not\!\partial-m)\Psi^i
                                    \nonumber \\
 L_{int}&=& -i\lambda\frac{\sum_i
  (\overline{\Psi}^{i}\mbox{\boldmath $\tau$}\gamma_{\mu}\Psi^i)\cdot\sum_j   
 (\overline{\Psi}^j\mbox{\boldmath $\tau$}\stackrel{\leftrightarrow}{\partial}^{\mu}\Psi^j)}{
  2\sum_k (\overline{\Psi}^k\Psi^k)},\;\;\;(\lambda\rightarrow 1).        \label{NAf}
\end{eqnarray}
Gauge invariance can be proved in a similaar way as in the bosonic model although the 
auxiliary field method never leads us to this Lagrangian.  To the first order 
in $\mbox{\boldmath $\alpha$}(x)$ under the space-time dependent rotation 
$\Psi\rightarrow \exp(i\mbox{\boldmath$\tau$}\cdot\mbox{\boldmath $\alpha$(x)}/2)\Psi$ 
and its conjugate, the gauge variations are given by
\begin{eqnarray}
 \delta L_0 &=& -\frac{1}{2}(\overline{\Psi}\gamma_{\mu}\mbox{\boldmath $\tau$}\Psi)
             \cdot \partial^{\mu}\mbox{\boldmath $\alpha$} + O(\alpha^2), \nonumber \\
 \delta L_{int} &=& -\lambda\delta L_0.    \;\;\; (\lambda \rightarrow 1)   \label{varf} 
\end{eqnarray}
We can prove the gauge invariance to all orders of $\mbox{\boldmath$\alpha$}(x)$ using
Eq.(\ref{finite}).  In fact, a brute-force proof to all orders of $\alpha$ is 
mathematically less cumbersome for the fermionic model than for the bosonic model.

Just as in the case of bosonic matters, this simple form of the non-Abelian model 
is possible only for the doublet matter fields in SU(2) gauge symmetry.  It should 
be emphasized that our non-Abelian fermionic model cannot be obtained from the 
Lagrangian of nonpropagating auxiliary vector fields. 
   
\subsection{Noether current}
As it happens in the Abelian models, the Noether current does not exist in 
our bosonic nor fermionic non-Abelian models. The reason is the same as 
in the Abelian case: For the Lagrangians with the matter fields alone, 
the contributions to the Noether current from $L_0$ and $L_{int}$ cancel 
each other as a very consequence of gauge invariance. The proof in
Appendix A can be trivially extended to the non-Abelian models. Even without
such a general proof, the Noether currents off the gauge symmetry limit, 
which are given below, clearly show their absence in the gauge symmetry limit.  

The Noether current exists off the gauge symmetry limit. Following the standard
prescription, we obtain the Noether currents from our Lagrangians of
Eqs.(\ref{NAb}) and (\ref{NAf}) in the form,
\begin{eqnarray}
     \mbox{\bf J}^N_{\mu}&= &
        i(1-\lambda)\Phi^{\dagger}\frac{\mbox{\boldmath $\tau$}}{2}
  \stackrel{\leftrightarrow}{\partial}_{\mu}\Phi , \;\;(\mbox{bosonic})\nonumber\\
     \mbox{\bf J}^N_{\mu}&= &
  (1-\lambda)\overline{\Psi}\frac{\mbox{\boldmath $\tau$}}{2}\gamma_{\mu}\Psi .
                               \;\;   (\mbox{fermionic}) 
\end{eqnarray} 

As for the energy-momentum tensor, the conserved tensor operator exists for 
any value of $\lambda$ just as in the U(1) models. 

\subsection{Composite gauge bosons}

In the case of the SU(2)-doublet matter fields, the non-Abelian diagram calculation 
is almost identical with the Abelian one. The only difference is in the insertion 
of the $\mbox{\boldmath $\tau$}$ matrix at every point of vectorial interactions 
in Fig. 1 and Fig. 7.  The massless composite bosons emerge in the $J^{PC} = 1^{--}$ 
channels of the adjoint representation of SU(2). In the case of fermion matter 
the composite massless bosons appear in the $^3S_1$ eigenchannel, 
that is, they couple only through $\overline{\Psi}\mbox{\boldmath $\tau$}
\gamma_{\mu}\Psi$.  The correct properties of the massless bound states are 
confirmed just as in the Abelian cases.

 We summarize the difference of the SU(2)-doublet models from the Abelian models:

(A) For the non-Abelian models of SU(2)-doublet matter fields, the vacuum expectation 
value $I_0^b =\langle 0|\Phi^{\dagger}\Phi|0\rangle$ and 
$I_0^f=\langle 0|\overline{\Psi}\Psi|0\rangle$ are twice as large as their 
Abelian values, respectively, since both the upper and lower components of 
the doublet matter contribute. 

(B) The bubble diagrams entering the kernel $K$ of the iteration equation are 
scaled upward by the same factor of two since a trace is taken within the 
bubble loop; ${\rm tr}(\tau_a\cdot\tau_b) = 2\delta_{ab}$. 

(C) Since the multiplication of the factor two in (A) and (B) occurs in 
both the numerator and the denominator of the kernel $K$ in Eq.(\ref{kernel1}) 
and Eq.(\ref{Kelement}), it keeps the kernel $K$ unchanged from the Abelian 
value. Meanwhile, the lowest-order T-matrix, $T^0$, is scaled down by factor 
two since it is inversely proportional to $I_0^b$ ($I_0^f$).  So is the 
amplitude $T=(I-K)^{-1}T_0$.   

Since the kernel $K^{\mu\nu}$ remains unchanged, $(I-K)$ is still transverse 
and starts with a term proportional to $g^{\mu\nu}q^2-q^{\mu}q^{\nu}$ with 
the same nonvanishing coefficient. Consequently the solution for the iterated 
amplitude $T$ takes the same form as in the corresponding Abelian models, 
but the residue at $q^2 =0$ is half as large, reflecting the fact that the 
lowest-order term $T^0$ is scaled down by factor two.  

Summing up  
this argument, the location of the pole at $q^2=0$ remains the same and 
its residue is scaled down by factor two, relative to the Abelian models,
for both the bosonic and the fermionic model. We describe below some more
details specific to each of the non-Abelian models.
 
\vskip 0.5cm
\underline{\bf The bosonic model}

We compute the chain of bubble diagrams as shown in Fig. 1 where the 
$\mbox{\boldmath$\tau$}$-matrices are inserted at every point of 
interaction.  The residue at the massless pole is compared with that 
of the corresponding Feynman diagram computed with the standard 
Lagrangian of the SU(2) gauge symmetry,
\begin{equation}
  L_{int}=ig_2\Bigl(\Phi^{\dagger}\mbox{\boldmath $A$}^{\mu}\partial_{\mu}\Phi
               -\partial_{\mu}\Phi^{\dagger}\mbox{\boldmath $A$}^{\mu}\Phi \Bigr)
               +g_2^2\Phi^{\dagger}(\mbox{\boldmath $A$}^{\mu}\cdot
                        \mbox{\boldmath $A$}_{\mu})\Phi,   \label{effLb}
\end{equation}
where ${\bf A}^{\mu} = \frac{1}{2}\tau_a A_a^{\mu}$. We obtain the gauge coupling
of the composite SU(2) gauge bosons ${\bf A}^{\mu}$ to the matter fields, 
\begin{equation}
       \frac{g^2_2}{4\pi} =  \frac{96\pi^2}{N\ln(\overline{\Lambda}^2/m^2)},
                                  \label{NAbcoupling}
\end{equation}
when it is expressed with the cutoff $\overline{\Lambda}$ in the space-time 
dimension of four.\footnote{For $\overline{\Lambda}$, see 
Eq.(\ref{U1bcoupling1}) and the line following it.}  Recall that the standard 
definition of $g_2$ accompanies the generators $\frac{1}{2}\mbox{\boldmath $\tau$}$ 
instead of just $\mbox{\boldmath $\tau$}$.  (See the definition of 
${\bf A}^{\mu}$ following Eq.(\ref{effLb}).  In the leading 1/N order, the 
magnitude of coupling Eq.(\ref{NAbcoupling}) coincides with what one would 
obtain in the auxiliary field trick since it comes from the same single 
bubble diagram with $\mbox{\boldmath$\tau$}$ on the both ends.

 The four-body interaction 
$\Phi^{\dagger}\Phi \mbox{\boldmath A}_{\mu}\mbox{\boldmath A}^{\mu}$ 
can be computed with the second term of the expansion for
$1/(\Phi^{\dagger}\Phi)$ around its vacuum value in $L_{int}$, namely,
\begin{equation} 
  -\frac{1}{4(I_0^b)^2}(\Phi^{\dagger}\mbox{\boldmath $\tau$}\stackrel{\leftrightarrow}{
\partial}_{\mu}\Phi)\cdot(\Phi^{\dagger}\mbox{\boldmath $\tau$}\stackrel{\leftrightarrow}{
\partial}^{\mu}\Phi)(:\Phi^{\dagger}\Phi:).         
\end{equation}
Attaching chains of bubbles to $(\Phi^{\dagger} \mbox{\boldmath $\tau$}\stackrel{
\leftrightarrow}{\partial}^{\mu}\Phi)$ and $(\Phi^{\dagger} \mbox{\boldmath $\tau$}
\stackrel{\leftrightarrow}{\partial}_{\mu}\Phi)$ of this interaction and approaching 
the zero momentum limit, we obtain $g_2^4$, of which $g^2_2$ is assigned to the gauge 
couplings of two composite gauge bosons with the external 
$\Phi^{\dagger}\mbox{\boldmath $\tau$}\Phi$ at the outer ends of the chains and 
the remaining $g_2^2$ is assigned to the $\Phi^{\dagger}\Phi A_{\mu}A^{\mu}$ coupling.  
This step is a repeat of what we have done for the Abelian model depicted in Fig. 3
and Fig. 4.  Going through this computation, we find that the resulting $g^2_2$ 
for $\Phi^{\dagger}(\mbox{\boldmath $A$}^{\mu}\cdot\mbox{\boldmath $A$}_{\mu})\Phi$ 
is equal to the value given in Eq.(\ref{NAbcoupling}), as we expect.

For the non-Abelian gauge bosons, there must be the triple self-coupling and the 
quartic self-coupling. They are computed with the diagrams of Fig.5 and Fig.6 
after inserting the $\tau$-matrices appropriately. The triple self-coupling 
diagrams, of course, do not cancel among themselves in the non-Abelian case. 
Charge conjugation invariance allows the triple self-coupling since the non-Abelian 
charge flowing in the opposite directions in a pair of triangular diagrams survives 
with $\tau_a\tau_b-\tau_b\tau_a = 2i\epsilon_{abc}\tau_c\neq 0$. Paying attention 
to the subtlety of the linear divergence that has been cautioned earlier, 
we find that the value obtained for the triple self-coupling agrees with what the 
SU(2) gauge symmetry requires by $-\frac{1}{4}{\bf G}_{\mu\nu}\cdot{\bf G}^{\mu\nu}$.  
The quartic self-coupling arises from the diagrams with four-corner, three-corner 
and two-corner loops at the center (i.e., Fig. 6a, 6b, and 6c) and survives 
in the limit of zero external momenta. They have the correct magnitude and 
group structure as required by the SU(2) gauge symmetry.

 All this should not be surprising after we have found a triplet of spin-one 
massless bound states out of the manifestly gauge invariant Lagrangian.  
Once we have found that the effective fields of these bound states couple 
with the matter fields in the form
\begin{equation}
     L_{int} = ig_2(\Phi^{\dagger}\mbox{\bf A}^{\mu}\partial_{\mu}\Phi
              -\partial_{\mu}\Phi^{\dagger}\mbox{\bf A}^{\mu}\Phi), \label{Yc}
\end{equation}
with ${\bf A}^{\mu}=\frac{1}{2}\tau_aA_a^{\mu}$, all other couplings of 
${\bf A}_{\mu}$ necessary to satisfy the SU(2) gauge invariance ought to 
be generated by loop and chain diagrams in the same 1/N order.  Otherwise 
the models would violate the SU(2) gauge invariance that has been embedded 
in Lagrangian at the beginning. We know no other way to be compatible with 
the SU(2) gauge symmetry once the interaction of Eq.(\ref{Yc}) emerges. 

\vskip 0.5cm
\underline{\bf The fermionic model}    

Let us turn to the fermionic model.
While presence of two $J^{PC}=1^{--}$ channels requires $2\times 2$ matrix 
calculation, the diagram computation of the bound-state generation is identical  
with that of the Abelian case except for insertion of the $\mbox{\boldmath$\tau$}$
matrices into the $2\times 2$ matrix equation of Fig. 7 after replacing the boson 
lines with the fermion lines.  Massless bound states 
appear in the $^3S_1$ channel here again and the squared SU(2) gauge coupling 
expressed in $g^2_2$ is larger than that of the U(1) fermionic model by factor two 
just as in the bosonic case:
\begin{equation}
    \frac{g_2^2}{4\pi} = \frac{24\pi^2}{N\ln(\overline{\Lambda}^2/m^2)}, 
                              \label{NAfcoupling}
\end{equation}
where the coupling $g_2$ is defined by
\begin{equation}
   L_{int}= -g_2\overline{\Psi}\gamma_{\mu}\mbox{\bf A}^{\mu}\Psi. 
\end{equation}
When we work on the other couplings of dimension four, we do not encounter
any complication new to the non-Abelian symmetry.  The reason is that the 
massless bound states couple to the matter fields only through the vertex of 
$(\overline{\Psi}\gamma_{\mu}\mbox{\boldmath$\tau$}\Psi)$, 
not through $(\overline{\Psi}\mbox{\boldmath$\tau$}
\stackrel{\leftrightarrow}{\partial}_{\mu}\Psi)$.
Therefore the computation of the triple and quartic self-couplings
can be carried out in the same way as in the U(1) model.   
The relevant diagrams are those of Fig. 5 and Fig. 6 where the boson lines 
are replaced with the fermion lines.  Since the composite bound states 
generated in the chains of bubbles couple with the fermions only through
$(\overline{\Psi}\mbox{\boldmath $\tau$}\gamma_{\mu}\Psi)$, not through 
 $(\overline{\Psi}\mbox{\boldmath $\tau$}
\stackrel{\leftrightarrow}{\partial}^{\mu}\Psi)$, the vertices of the 
triangle (Fig. 5) and the box (Fig. 6a) at the center of diagram are only 
those of $\gamma_{\mu}$, not of $\stackrel{\leftrightarrow}{\partial}_{\mu}$.
The diagrams of Fig. 6b and Fig. 6c do not contribute since the six-body 
interaction 
$(\overline{\Psi}\mbox{\boldmath $\tau$}\gamma_{\mu}\Psi)(\overline{\Psi}
\mbox{\boldmath $\tau$}\stackrel{\leftrightarrow}{\partial}^{\mu}\Psi)
(\overline{\Psi}\Psi)$
is incapable of producing two composite bosons. (Recall the argument in 
the Abelian fermionic model.) As for the fermionic triangular and box
diagrams corresponding to Figs. 5 and 6a, the same large-$N$ computation 
was actually carried out twenty years ago in a similar model\cite{MS} 
that contains an explicit gauge-symmetry breaking but only through 
the gauge boson mass.  We do not repeat the calculation of the 
self-couplings for the non-Abelian fermionic model here. The bottom line 
is that the same coupling $g_2$ as the matter-gauge-boson coupling 
of Eq.(\ref{NAfcoupling}) appears in the self-interaction of 
the gauge bosons as we expect. 

All these beautiful outcomes conforming to non-Abelian gauge symmetry are 
manifestation of gauge invariance that is embedded in the Lagrangian at 
the beginning. Hoping that we are not overly repetitious, we emphasize 
that once the massless bound states of spin-one appear and their effective
fields ${\bf A}^{\mu}$ couple with the matter fields like 
$g_2\overline{\Psi}\gamma_{\mu}{\bf A}^{\mu}\Psi$, the bound states must 
be gauge bosons and the associated gauge self-couplings of ${\bf A}^{\mu}$ 
in $-\frac{1}{4}{\bf G}_{\mu\nu}{\bf G}^{\mu\nu}$ must be generated in order 
to satisfy SU(2) gauge invariance of $L_{tot}$.  We know no other way to 
realize the non-Abelian gauge invariance.
 
\section{Discussion}

We start the final section with an obvious observation common to all of 
our models. In our models we cannot introduce an elementary gauge field 
by the method of the substitution rule
$\partial_{\mu}\rightarrow \partial_{\mu}+ieA_{\mu}$ in our Lagrangian.
The reason is obvious by the structure of the models: This substitution 
operation is nothing other than one special gauge transformation.  Take 
for example the fermion fields $\psi$ in our U(1) Lagrangian. The 
substitution $\partial_{\mu}\psi\rightarrow (\partial_{\mu}+ieA_{\mu})\psi$ 
is realized by the rotation 
\begin{equation}
   \psi(x) \rightarrow \exp\bigl(ie\int^x A_{\mu}(y)dy^{\mu}\bigr)\psi(x).  
                        \label{dis1} 
\end{equation}
Since Eq.(\ref{dis1}) is one of the gauge transformations with
\begin{equation}
   \alpha(x) = e\int^x A_{\mu}(y)dy^{\mu},   \label{phase}
\end{equation}
the function $\alpha(x)$ is canceled out between $L_0$ and $L_{int}$ by gauge 
invariance and disappears from Lagrangian entirely. Therefore the elementary
$A_{\mu}$ field cannot be introduced into our Lagrangians in this way.  
Inability to introduce the elementary $A_{\mu}$ field in our Lagrangians 
by the so-called substitution rule is closely in parallel with vanishing 
of the Noether current.  

  The next observation concerns the no-go theorem of Weinberg and Witten.  
The theorem was stated in the following way \cite{WW}:

\underline{Theorem} {\em A theory that allows the construction of a Lorentz-covariant 
conserved four-vector current $J^{\mu}$ cannot contain massless particles of spin 
$j > 1/2 $ with nonvanishing values of the conserved charge $\int J^0 d^3x.$} 

 The proof is simple. Fix first the Lorentz scalar value of the matrix element 
$\langle p'|J_{\mu}|p\rangle$ for the massless spin-one particle in the forward 
limit $p'\rightarrow p$. Then make a Lorentz transformation and examine its 
rotational property around the momentum ${\bf p}$ in the brick-wall frame 
(${\bf p}' = - {\bf p}$).  We need the conserved current $J_{\mu}$ that provides 
the Lorentz scalar charge $\int J_0d^3x$.     

The theorem holds whether the massless boson is elementary or composite. As was 
emphasized by the authors\cite{WW}, however, the theorem does not apply 
to the standard non-Abelian gauge bosons (without spontaneous symmetry breaking).  
The catch is in the word ``Lorentz-covariant''.  The state of zero helicity does not 
exist for massless gauge bosons.  In order to make the theory manifestly 
Lorentz covariant and gauge invariant at the same time, one has to fix a gauge 
by introducing an unphysical ghost state in the Lagrangian.  Otherwise, one cannot 
carry out diagram calculation.  Fixing a gauge by a subsidiary condition either 
violates manifest gauge invariance or introduces a state that does not 
exist as a physical particle state. Therefore, Lorentz scalar charges that meet 
the conditions of the Theorem do not exist in the standard non-Abelian gauge 
theory.\footnote{If one takes the purist 
viewpoint that the initial and final states of the matrix element 
$\langle p'|J_{\mu}|p\rangle$ must be asymptotic states, the theorem does not 
apply to the non-Abelian gauge theory like QCD, which is singular in the 
infrared limit so that one-gluon states are not asymptotic states.
Our non-Abelian models contain $N (\rightarrow \infty)$ doublets of matter 
particles so that the infrared limit is nonsingular, i.e., not confining.} 

  What should we do with this theorem for our non-Abelian models ? If we could 
write the non-Abelian Noether currents with the matter fields alone, we would 
potentially interfere with this theorem. However, the Lorentz-covariant conserved
currents do not exist in our models. They exist only off the gauge symmetry 
limit ($\lambda \neq 1$) and disappear as we go to the gauge symmetry limit of 
$\lambda = 1$, and it is only at this point that the vector bound states 
become massless.  We thus circumvent the theorem. Is this really the answer 
to the potential conflict of the composite non-Abelian gauge bosons 
with the Weinberg-Witten theorem ?  To be frank, the present author is not 
totally comfortable with this answer.  But it appears in our models that 
generation of the massless non-Abelian composite bosons evades the conflict 
with the Weinberg-Witten theorem.\footnote{The W and Z bosons 
in the extra dimension model\cite{Gher} are the lowest lying 
Kaluza-Klein modes with mass so that they do not conflict with the theorem.}

   It is explicitly visible in our models that gauge invariance requires 
that the force in the $1^{--}$ channel be attractive ($\lambda > 0$) 
and that the bound state in this channel be massless $(\lambda\to 1$ ).
Repulsive forces ($\lambda < 0$) cannot be gauge invariant.
We are tempted to speculate that {\em even if gauge fields are not 
introduced explicitly, gauge bosons must appear as composite states 
if a theory is gauge invariant.}  While it sounds like a trivial 
proposition, it is desirable to elevate it to a rigorous theorem 
of field theory. 

One obvious question is whether our models have anything to do with the real 
world.  At an early stage of the electroweak theory, people discussed the
possibility of composite W and Z.\cite{CFJ,ISL}  A quarter century ago the 
present author also joined to propose an unrenormalizable phenomenological
model of composite $W$ and $Z$ bosons which an explicit symmetry breaking 
enters only through the W/Z masses \cite{MS,G}. It was the time right 
after the experimental confirmation of the $W$ and $Z$ bosons by 
accelerator\cite{UA1,UA2}.  At that time very little was known 
experimentally about the properties of $W$ and $Z$. One sensitive 
theoretical test was to study how much deviation from the gauge symmetry 
could be accommodated for the self-couplings of dimension four through 
their loop contributions\cite{PL}.  More general test irrespective of 
sources was proposed \cite{PT} and is still being used for experimental 
test of the minimal standard model.  Now the Higgs boson has been 
discovered with its properties roughly in agreement with the theoretical 
expectation, the next step is to raise precision in the interaction of 
$W$ and $Z$ by direct measurement. The early indication of the two-photon
anomaly at 750GeV is one example that may open up a new window. However,
since the invariant mass of 750GeV is near the upper end of the two-photon 
phase space in the current data and ``the anomaly'' is still no more than 
a three-standard-deviation effect even with the ATLAS and CMS data 
combined, we need to wait some time before a consensus is reached among 
experimentalists on this anomaly.  Both experimentalists and phenomenologists 
are working toward to this goal \cite{Atlas,Alioli}.

When our model is expressed as a composite gauge theory with the 
effective fields ${\bf A}_{\mu}$, difference from the minimal standard
model would appear in the interactions of dimension higher than four 
which are suppressed by powers of 
$p^2/\overline{\Lambda}^2$ at $|p^2| < \overline{\Lambda}^2$. 
When experiment explores the region of energies comparable or higher 
than $\overline{\Lambda}$, shall we be able to discriminate directly
our model Lagrangian from the standard model of $W$ and $Z$. But we
have no theoretical basis to speculate on magnitude of 
$\overline{\Lambda}$ at present. 

We conclude with one disturbing question to which we give no good answer. 
Is it really possible to tell experimentally or even theoretically whether 
a given particle is elementary or composite ?  This is a nagging question 
that confronted theorists\cite{N} at the height of nuclear democracy in 
the early 1960's. Theorists proposed various criteria of compositeness, 
but no consensus emerged. Although we have started with the matter fields 
alone and constructed the massless gauge bosons explicitly as their 
bound states, can't we describe exactly the same physics with some other
Lagrangian in which all particles are elementary ? Can we really answer 
the question of elementarity vs compositeness once for all ?  

The following theorem was given by Kamefuchi, O'Rafeartaigh and 
Salam\cite{KOS} in 1961: If a composite local operator carries all quantum 
numbers of a given particle in regard to space-time ($J^{PC}$) and other 
properties (charge, isospin etc), it gives the same S matrix amplitudes on 
the particle mass shell up to overall normalizations. Difference shows up 
only off the mass shell. But the ``off-shell amplitudes'' are not really 
scattering amplitudes of the particle, but include continuum contributions.  
According to this theorem, therefore, the definition of particle fields 
is infinitely ambiguous with respect to their continua. When a different 
particle field is used, its interaction Lagrangian takes a different form. 
To avoid this ambiguity and the issue of renormalizability, we were tempted 
to replace the field theory with the S-matrix theory in the 1960's so as 
to deal only with the on-shell amplitudes and the observables. 
As we know, it led us to the dual resonance model and then back to 
Lagrangian theory of strings with the Nambu-Goto action. 

Meanwhile, the present author has been brought attention to one interesting 
observation in supersymmetric theory. Along the line of the Olive-Montonen 
conjecture, Seiberg and Witten\cite{SW} showed in the $N=2$ supersymmetric 
theory that the strong and weak coupling limits are dual to each other. 
To be more specific, the roles of a particle and a soliton of the same 
spin-parity are interchanged between the strong and weak limits of coupling. 
Since solitons are composite in everyone's picture, in such theories 
elementarity vs compositeness loses its absolute meaning. It depends on 
the strength of coupling. The similar duality was shown earlier for 
a model of N=4 too.\cite{Osborn} 
  Proof of this duality relies on the simple holomorficity special to 
supersymmetry. If something similar holds in nonsupersymmetric theory as
well, the meaning of elementarity and compositeness of particles would 
finally disappear and the naming would become just for a matter of 
convenience; if Lagrangian takes the simplest form with a certain choice 
for a set of particle fields, one would call such particles as 
{\em elementary} for convenience.  

\appendix
\section{Nonexistence of Noether current}

The Noether current does not exist in the theories that satisfy local
gauge invariance with matter fields alone.  The proof is almost trivial.
We give it here only for the U(1) bosonic model since extension to 
fermions and non-Abelian theories is straightforward.

Under the U(1) gauge transformation, the Lagrangian satisfies the local 
invariance,
\begin{equation}
   L(e^{-i\alpha(x)}\phi^*,e^{i\alpha(x)}\phi) = L(\phi^*,\phi),     \label{Ag}
\end{equation}
where $\alpha(x)$ is an arbitrary function of space-time that satisfies mild conditions 
such as differentiability. The copy index $i$ ($=1,\cdots N$) has been suppressed
in Eq.(\ref{Ag}).  For the infinitesimal $\alpha(x)$, gauge invariance requires
\begin{eqnarray}
    -i\Bigl(\phi^*\frac{\partial L}{\partial\phi^*}+
     \partial_{\mu}\phi^*\frac{\partial L}{\partial(\partial_{\mu}\phi^*)}\Bigr)\alpha
  &+&i\Bigl(\frac{\partial L}{\partial\phi}\phi +
      \frac{\partial L}{\partial(\partial_{\mu}\phi)}\partial_{\mu}\phi\Bigr)\alpha
           \nonumber \\
  &+&i\Bigl(-\phi^*\frac{\partial L}{\partial(\partial_{\mu}\phi^*)}
    +\frac{\partial L}{\partial(\partial_{\mu}\phi)}\phi\Bigr)\partial_{\mu}\alpha =0.
           \label{Agg}
\end{eqnarray}  
Since $\alpha(x)$ and $\partial_{\mu}\alpha(x)$ are two independent functions when 
$\alpha(x)$ is an arbitrary function of $x_{\mu}$, the condition of Eq.(\ref{Agg}) 
requires that the terms proportional to $\alpha(x)$ and to $\partial_{\mu}\alpha(x)$ 
must be separately equal to zero.  After use of the equations of motion, the 
coefficient of $\alpha(x)$ equal zero gives
\begin{equation}
     -\partial_{\mu}\Bigl(\phi^*\frac{\partial L}{\partial\phi^*_{\mu}}\Bigr)
 +\partial_{\mu}\Bigl(\phi\frac{\partial L}{\partial\phi_{\mu}}\Bigr)
               = 0. \label{DivNoether}
\end{equation}
Normally this would be the statement of conservation of the Noether current, 
$\partial^{\mu}J_{\mu}^N=0$.  However, the third term proportional to 
$\partial_{\mu}\alpha(x)$ in Eq.(\ref{Agg}) gives
\begin{equation}
 -\frac{\partial L}{\partial(\partial_{\mu}\phi^*)}\phi^*
         +\frac{\partial L}{\partial(\partial_{\mu}\phi)}\phi = 0. \label{Aggg}
\end{equation}
This is nothing other than the statement of
\begin{equation}
                   J_{\mu}^N \equiv 0                \label{Acurrent}
\end{equation}
at all space-time locations. In the case that the elementary gauge field 
$A_{\mu}$ exists in Lagrangian, the gauge transformation 
$A_{\mu} \rightarrow A_{\mu}+i\partial_{\mu}\alpha$ generates an additional 
term proportional to $\partial_{\mu}\alpha(x)$ and adds to the third term 
in Eq.(\ref{Agg}) to cancel exactly the variation due to $\phi/\phi^*$. This 
cancellation is nothing other than gauge invariance itself.  Consequently,
Eq.(\ref{Acurrent}) does not follow in the conventional gauge theory.
Extension of this proof to the fermion models and the non-Abelian models 
is just as simple and easy.

  Despite this general proof of $J_{\mu}^N \equiv 0$, some may wonder if it is 
possible to define a conserved current in the gauge symmetry limit by factoring 
out $(1-\lambda)$ from the current $J_{\mu}$ defined by Eq.(\ref{conservedb}) 
off the gauge limit ($\lambda \neq 1$) and then going to the limit of 
$\lambda = 1$.  If physics is somehow ``continuous'' in this respect in the 
neighborhood of $\lambda=1$, this might allow us to circumvent the difficulty.  
That is, choose as a conserved current simply the current
\begin{equation}
 J'_{\mu}  = i\sum_i(\phi_i^*\stackrel{\leftrightarrow}{\partial}_{\mu}
                                \phi_i),  \label{conb}
\end{equation}
so that the charge is $Q \equiv \int J'_0d^3x$.  This charge is not gauge
invariant, but let us leave it aside for a moment.
If one computes by brute force the divergence of this current $J'_{\mu}$ 
with the equation of motion, one would not be led to
$\partial^{\mu}J'_{\mu} = 0$.
Instead one would end up with the trivial circular identity as follows: 
Since $\partial^{\mu}J'_{\mu} = i\sum_i(\phi_i^*\Box\phi_i -\Box\phi_i^*\phi_i)$, 
one multiplies the equation of motion for $\phi_i$ with the field $\phi^*_i$ and 
subtracts the corresponding bilinear object with $\phi_i\leftrightarrow \phi^*_i$. 
Then the result is a trivial identity: $i\sum_i(\phi_i^*\Box\phi_i -
\phi_i^*\Box\phi_i)=i\sum_i(\phi_i^*\Box\phi_i -\Box\phi_i^*\phi_i)$. 
Therefore the conclusion from this exercise is as follows:
Only when one violates gauge invariance by staying away from the symmetry 
limit ($\lambda\neq 1$), can the Noether theorem define a conserved current 
in the familiar form with strength reduced by $(1-\lambda)$. 

   The same happens for our fermion model.  Just as in the bosonic model, 
the current $\sum_i\overline{\psi}_i\gamma_{\mu}\psi_i$ is not the conserved 
Noether current in the gauge symmetry limit.\footnote{Unlike the corresponding 
object in the bosonic case, this current is gauge invariant.}  
The equation of motion of $L_{tot}$ does not allow us to compute 
$\partial^{\mu}(\overline{\psi}\gamma_{\mu}\psi)$ in the gauge symmetry limit:
Such computation drives us around a circular loop just as in the case of bosons.

In the perturbative diagram calculation which is performed in the interaction
picture, however, the fields obey the equation of {\em free} motion. 
Therefore $\phi^*\stackrel{\leftrightarrow}{\partial}_{\mu}\phi$ 
and $\overline{\psi}\gamma_{\mu}\psi$ are both divergence free, that is, 
conserved currents.

\section{Difficulty in general non-Abelian models}

The local Lagrangian of matter fields alone has been easily obtained by the 
auxiliary gauge fields method for the SU(2) model with the doublet matter. 
But we cannot extend it to other groups and representations.  We show it here
with two explicit examples.

Let us start with the Lagrangian of the nonpropagating auxiliary gauge fields,
\begin{equation}
   L= \Phi^{\dagger}(\stackrel{\leftarrow}{\partial}^{\mu}-i{\bf A}^{\mu})
         (\partial_{\mu}+i{\bf A}_{\mu})\Phi - m^2\Phi^{\dagger}\Phi
         +\frac{1}{2}\mu^2A_{a,\mu}A^{\mu}_a, \;\;(\mu^2\rightarrow 0)
                       \label{BL1}
\end{equation}  
where $\Phi$ and $\Phi^{\dagger}$ are the column and row fields belonging to the 
$n$-dimensional representation of group $G$.  We have absorbed the coupling $e$
into ${\bf A}_{\mu}$.  Let the group $G$ be induced by 
the generators $T_a$ ($a= 1,\cdots k$), which are $n\times n$ matrices.  
We represent the nonpropagating gauge fields $A^{\mu}_a (a=1\cdots k)$ in 
the $n\times n$ matrices,
\begin{equation} 
        {\bf A}^{\mu} =  \sum_{a=1}^kT_aA_a^{\mu}. \label{BA2}
\end{equation} 
The Lagrangian Eq.(\ref{BL1}) is invariant under the local gauge transformation,
\begin{eqnarray}
     \Phi &\rightarrow& U\Phi, \nonumber \\
   A^{\mu} &\rightarrow& U A^{\mu}U^{\dagger}-i(\partial^{\mu} U)U^{\dagger}.  \label{Bgt}
\end{eqnarray}
where $U = \exp(iT_a\alpha_a)$.
In order to integrate the exponentiated action of $L$ over $A^{\mu}_a$, 
we combine the terms bilinear and linear in $A^{\mu}_a$ into a quadrature 
and ``shift the origin''. In the case of the SU(2)-doublet matter fields, we 
see with $\{\tau_a,\tau_b\}=2\delta_{ab}$ that the coefficients of 
the bilinear terms of $A^{\mu}_a$ are simply $\delta_{ab}\Phi^{\dagger}\Phi$ 
so that no diagonalization is needed for symmetrized product of the generators 
$\{T_a,T_b\}=\frac{1}{4}\{\tau_a,\tau_b\}=\frac{1}{2}\delta_{ab}$. Upon integration
over $A^{\mu}_a$, the denominator of $L_{int}(\Phi^{\dagger},\Phi)$ comes out to 
be the singlet $\Phi^{\dagger}\Phi$, as given in Eq.(\ref{NAb}). Upon integration, 
an additional term
\begin{equation}
         -2{\rm tr}\ln(\Phi^{\dagger}\Phi)
\end{equation}  
appears in the effective action.  But we may remove this term since it is 
gauge-invariant by itself.  We retain the remainder as the gauge-invariant 
Lagrangian in terms of $\Phi^{\dagger}/\Phi$. 

However, this procedure does not work in the cases other than the SU(2) doublet. 
When $\{T_a,T_b\}\not\propto\delta_{ab}I$, it happens that the integral over 
${\bf A}_{\mu}$ is generally impossible.  Even if it were possible,
the {\em trace-log} term would not be invariant by itself under rotations 
of group G, not even under global rotations. While the whole action is 
gauge invariant, it is not separately so for the effective Lagrangian and 
the {\em trace-log} term.  Unfortunately, this is what happens in the
cases other than the SU(2) doublet.  We show two simple examples below. 
 
Let us first examine the case of the real triplets of SU(2). 
In this case the coefficient of the bilinear terms of $A_a^{\mu}$ ($a=1,2,3$) is 
written in terms of the $3\times 3$ matrices $(T_a)_{bc}=-i\varepsilon_{abc}$ and the 
matter fields $\Phi = (\phi_1,\phi_2,\phi_3)^t$ and $\Phi^{\dagger} = \Phi^t$.  
The bilinear terms of $A_a^{\mu}$ is given by
\begin{equation}
      (\Phi^tT_aT_b\Phi) A_a^{\mu}A_{b,\mu}. \label{bilinear}
\end{equation}
It can be diagonalized by the orthogonal transformation $A'_{\mu}= {\bf O}A_{\mu}$ into
\begin{equation}
  (A^{'\mu}_1, A^{'\mu}_2, A^{'\mu}_3) 
                       \left(  \begin{array}{ccc}
                       \Phi^t\Phi & 0 & 0 \\
                       0&  \Phi^t\Phi & 0 \\
                       0&  0 & 0  \end{array} \right) 
                   \left(  \begin{array}{c}
                                 A'_{1 \mu}\\
                                 A'_{2 \mu}\\
                                 A'_{3 \mu} 
                          \end{array} \right).
\end{equation}            
When this is placed in the action and exponentiated, we cannot integrate it 
over the third component of ${\bf A}'_{\mu}$ since the action is flat along 
that direction (at $\mu\rightarrow 0$). The action blows up as $\mu\rightarrow 0$
and there is no way to keep it well-defined. 

How about the SU(3)-triplet matter fields as the next-to-simplest example?  
For the triplet matter fields, the bilinear terms in $A_a^{\mu}(a=1,\cdots 8)$
can be written as 
\begin{equation}
                      A_{\mu}^a M_{ab} A^{b,\mu},
\end{equation}
where $M_{ab}=\frac{1}{8}\Phi^{\dagger}\{\lambda_a,\lambda_b\}_{+}\Phi$ is 
a symmetric matrix under $a\leftrightarrow b$. The matrix $M_{ab}$ can be 
diagonalized into ${\bf D}$ by some orthogonal rotation {\bf O} as
\begin{equation}
    (A'_{\mu})^t{\bf O}^t M {\bf O}A'^{\mu} =   A'_{a,\mu}D_{aa}A'^{\mu}_a.
\end{equation} 
Can the diagonal matrix ${\bf D}$ be proportional to the unit matrix? 
If so, the functional integral over $A_{\mu}^a$ would produce a denominator 
common to all $a$ in $L_{int}$ just as in the case of SU(2). But 
that is obviously not the case: If ${\bf D}\propto {\bf I}$, then 
$M_{ab}=({\bf O}{\bf D}{\bf O}^t)_{ab}$ would also have to be proportional to 
$\delta_{ab}$ even before the rotation.  We can easily see by simple inspection 
using the representation $T_a = \frac{1}{2}\lambda_a$ familiar to 
physicists, that $M_{ab}$ is not proportional to an $8\times 8$ unit matrix.
Consequently the resulting Lagrangian in terms of matter fields alone would 
not take a form as compact as in the SU(2) doublet case, if one could write 
it at all.\footnote{This does not conflict with what Rabinovici and Smolkin 
\cite{RS} did for general Lie groups: They integrate over the matter fields 
in one loop for a general group and representation to show that the 
$-\frac{1}{4}G_{\mu\nu}G^{\mu\nu}$ is indeed generated. Their purpose is to 
see whether or not this {\em Maxwell term} can be generated by loops upon 
integrating over matter fields in the auxiliary vector-field Lagrangian.  
They did not address to finding of a {\em local} non-Abelian gauge invariant 
Lagrangian written in matter fields alone.}  

These two examples show that the auxiliary field method can lead to a simple 
{\em local} field theory only for the U(1) and the SU(2)-doublet models of 
bosonic matter fields. 

\acknowledgments  

  I am thankful to Professor Pei-Ming Ho for bringing 
my attention to possible relevance or irrelevance of the auxiliary 
vector-field trick to compositeness of a gauge boson at an early stage.  
During the course of this work, I have been benefited with the useful 
conversation with Professor Korkut Bardakci.  After I proceeded 
substantially in this research, I learned from Professor Eliezer 
Rabinovic about some aspects of the subject which I had not been 
familiar with. Thanks also due to Professor E. Kh. Akhmedov for 
useful communications that directed me to many early works.

This work was supported by the Office of Science, Division of High Energy 
Physics, of the U.S.  Department of Energy under contract DE--AC02--05CH11231.


\newpage



\begin{thebibliography}{99}
\bibitem{YM}    C. N. Yang and R. L. Mills, Phys. Rev. {\bf 96}, 191 (1954).
\bibitem{CPN}   A. D'Adda, P. DiVecchia and M. Luscher, Nucl. Phys. B {\bf B146},
                63 (1978); {\em ibid} {\bf 152}, 125 (1979).
\bibitem{AV}    D. Amati and G. Veneziano, Nucl. Phys. {\bf B204}, 451 (1982).
\bibitem{Akh}   E. Kh. Akhmedov, Phys. Lett. B {\bf 521} 79 (2001).
\bibitem{Bj}   J. D. Bjorken, Ann. Phys. {\bf 24}, 174 (1963).
\bibitem{HHR}    H. Haber, I. Hinchliffe and E. Rabinovici, Nucl. Phys. {\bf B172}, 
                458 (1980).
\bibitem{Volk}   D. V, Volkov and V. P. Akulov, Pisma Zh. Eksp. Teor. Fiz. {\bf 16},
                 621 (1972): Phys. Lett. B {\bf 46}, 103 (1973).
\bibitem{Y}      H. Yukawa, Phys. Rev. {\bf 77}, 219 (1950).
\bibitem{Higgs}  P. Higgs, Phys. Rev. Lett. {\bf 12}, 132 (1964).
\bibitem{W}      S. Weinberg, Phys. Rev. Lett. {\bf 19}, 1264 (1967).
\bibitem{WW}     S. Weinberg and E. Witten, Phys. Lett. B {\bf 96}, 59 (1980).
\bibitem{BKY}    For review of earlier literature, see M. Bando, T. Kugo and K. Yamawaki, 
                 Phys. Rep. C {\bf 164}, 217 (1988) and references therein.
\bibitem{Raman}  R. Sundrum and L. Randall, Phys. Rev. Lett. {bf 83}, 3370 (1999). 
\bibitem{Gher}   B. Batell and T. Gherghetta, Phys. Rev. D {\bf 73}, 045016 (2006);
                 Y. Cui, T. Ghergetta and J. D. Wells, JHEP {\bf 0911}, 080 (2009); 
                 C. Csaki, Y. Shirman and J. Terning, Phys. Rev. D {\bf 84}, 095011 (2011). 
\bibitem{Rizzo}  H. Davoudiasl, J. L. Hewitt, and T. G. Rizzo, Phys. Lett. B {\bf 473}, 43 (2000);
                 T. Gherghetta and A. Pomarol, Nucl. Phys. {\bf B586}, 141 (2000);
                 K. Agashe, A. Delgado, M. J. May, and R. Sundrum, JHIP 08 (2003) 050.
\bibitem{RS}     E. Rabinovici and M. Smolkin, JHEP {\bf 1107}, 040 (2011).
\bibitem{BLS}    W. A. Bardeen, B. W. Lee and R. E. Shrock, Phys. Rev. {\bf D14},985 (1976).
\bibitem{CFJ}    M. Claudson, E. Fahri, and R. L. Jaffe, Phys. Rev. D {\bf 34}, 873 (1986).
\bibitem{ISL}    I-H. Lee and R. E. Shrock, Phys. Rev. Lett. {\bf 59}, 14 (1897);
                 Phys. Lett. {\bf 199B}, 541 (1987; {\bf 201}, 497 (1988);
                 S. Aoki, I-H. Lee, and R. E. Shrock, Phys. Lett. B {\bf 207}, 471 (1988).
\bibitem{MS}     M. Suzuki, Phys. Rev. {\bf D37}, 210 (1988).
\bibitem{G}      A. Cohen, H. Georgi and E. Simmons, Phys. Rev. D {\bf 38}, 405 (1988).
\bibitem{UA1}    UA1 Collaboration, Phys. Lett. {\bf 126B}, 398 (1983); {\em ibid} {\bf 134B}, 
                 469 (1984); {\em ibid} {\bf 166B}, 484 (1986).
\bibitem{UA2}    UA2 Collaboration, Phys. Lett. {\bf 129B}, 130 (1983); {\bf 186B}, 440 
                (1987).
\bibitem{PL}     M. Suzuki, Phys. Lett. {\bf 153B}, 289 (1985).
\bibitem{PT}    M. E. Peskin and T. Takeuchi, Phys. Rev. D {\bf 46}, 381 (1992).
\bibitem{Atlas}  The ATLAS Collaboration, ATL-PHY-PUB-2012-005.
\bibitem{Alioli} S. Alioli {\em et al}, CERN-TH-2016-137, CERN-LPCC-2016-002.
\bibitem{N}     B. Jovet, Nuovo Cim. {\bf 5}, 1133 (1956): M.T. Vaughn, R. Aaron,
                and R. D. Amado, Phys. Rev {\bf 124}, 1258 (1961); A. Salam, Nuovo Cim.
                {\bf 25}, 224 (1962); S. Weinberg, Phys. Rev {\bf 130}, 776 (1963),
                D. Lurie, A. J. MaCFarlane, Phys. Rev. {\bf 136}, B816 (1964);
                K. Akama and T. Hattori, Phys. Lett. {\bf 46B}, 106 (1998).
\bibitem{KOS}   S. Kamefuchi, L. O'Raifeartaigh, and A. Salam, Nucl. Phys. {\bf 28}, 529 (1961). 
\bibitem{SW}    N. Seiberg and E. Witten, Nucl. Phys. B {\bf 426}, 19 (1994).
\bibitem{Osborn} H. Osborn, Phys. Lett. B {\bf 83}, 321 (1979).
\end{thebibliography}
\end{document}